\documentclass[a4paper,11pt]{report}
\usepackage[utf8]{inputenc}
\usepackage[margin=2.5cm]{geometry}
\usepackage[titletoc,title]{appendix}
\usepackage{url}

\usepackage{setspace}
\usepackage{amsmath,amsfonts,amssymb,mathtools}
\usepackage{longtable}
\usepackage{graphicx,float,xcolor}
\usepackage[
backend=biber,
style=numeric,
citestyle=numeric
]{biblatex}
\usepackage{hyperref}
\hypersetup{colorlinks = true,
            linkcolor = blue,
            urlcolor  = blue,
            citecolor = red,
            anchorcolor = blue}

\title{\huge\textsc{{Robustness of excitations in the Random Dimer Model}}}
\author{ \Large Daniel Reti}
\date{\Large 9 September 2021}

\begin{document}

\maketitle
\begin{abstract}
    The ground state solution of the random dimer model is at a critical point after, which has been shown with random link excitations. In this paper we test the robustness of the random dimer model to the random link excitation by imposing the maximum weight excitation. We numerically compute the scaling exponents of the curves arising in the model as well as the fractal dimension. Although strong finite size corrections are present, the results are compatible with that of the random link excitation.
    Furthermore, another form of excitation, the $\epsilon$ - coupling excitation is studied. We find that near-optimal configurations belong to the same universality class as the travelling salesman problem. Thus, we confirm a conjecture on the scaling properties of combinatorial optimisation problems, for the specific case of minimum weight perfect matchings on 2-dimensional lattices. This document was submitted as my Thesis project for the MSc Complex Systems Modelling course at King's College London in 2021. In particular, I would like to give thanks to my supervisor, Dr Gabriele Sicuro for his insights and guidance.
\end{abstract}

\tableofcontents

\chapter{Introduction}

The field of statistical physics have tried classify models into universality classes to gain a better understanding of similarities in disordered systems. Universality implies that in the scaling limit when a large number of parts interact, systems are independent of dynamical features. Universality has been observed at critical systems by measuring critical scaling exponents

%why is it important

%critical systems description

%random dimer model is close to Ising model. Non-trivial ground state, 

Recently, the properties of the ground state solution of the random dimer model (RDM) was studied in \cite{caracciolo2021criticality}. More specifically, the authors showed that the ground state solution of the RDM is at a critical point. By the means of an accurate numerical algorithm, they showed that local excitations induce long-range rearrangement with finite probability. Furthermore, the fractal dimension of the excitations are computed. They find that the RDM on a bipartite graph is compatible with the loop-erased self-avoiding random walk, while the RDM on a monopartite lattice belongs to the same universality class 2-dimensional spin glass domain walls. 

The partial objective of this paper is to consider the RDM with maximum weight excitations and compare the results to those found in \cite{caracciolo2021criticality}, where the model was studied with the random link excitation. We are interested in how the curves generated with the maximum weight excitation behave in comparison to the random link excitation. The goal is to check the robustness of the RDM to a different form of excitation. Instead of uniformly sampling from the weight distribution of a given instance, we delete the edge with the largest weight in the ground state solution, with the expectation that that will cause the most "damage" to the optimal configuration. A large-scale Monte-Carlo simulation is undertaken with an exact algorithm to generate instances of the model. We find that the properties of the curves are in agreement with those found in \cite{caracciolo2021criticality}, although strong finite size corrections distort the numerical estimates. 

Furthermore, we examine the sensitivity of the RDM under a smooth excitation, namely the $\epsilon$ - coupling. The $\epsilon$ - excitation allows to infer the energy landscape of the RDM in close proximity to the ground state solution. Following the proposed conjecture on the universality of combinatorial optimisation problems in \cite{aldous2003scaling}, we study how the minimum weight perfect matching scales in the presence of $\epsilon$-couplings on 2-dimensional lattices. Similar scaling exponents are found to those reported in \cite{pagnani2003near}, although finite size corrections are present here too. The conjecture proposed in \cite{aldous2003scaling} appears to hold for the minimum weight perfect matching on lattices in 2-dimensions. 

In chapter \ref{ch.2}: \nameref{ch.2} we present a review of the relevant literature and key results. In chapter \ref{chap: max weight}: \nameref{chap: max weight} the maximum weight excitation of the RDM is analysed, including a discussion about connections to the random link excitation. In chapter \ref{ch.4}: \nameref{ch.4}, the study of the RDM under $\epsilon$ - coupling excitations is presented.

\chapter[Domino coverings and random assignments]{Domino coverings \\and random assignments} \label{ch.2}

%improve definitions here
% inf temp not defined here
The dimer model is part of a family of problems studied in matching theory. It consists in studying the properties of perfect matchings on a graph. A perfect matching of a graph is a subset of its edges that covers every vertex exactly once. The underlying graph structure can be monopartite or bipartite (or more). In the former case the problem is often referred to as the "matching problem", in the latter as the "assignment problem". Given a certain graph $G$, questions may arise about its the dimer coverings.For example, how many possible perfect matchings exist on $G$? Is there a way to count them? What are the properties of such matchings? 

%introduce weights in this paragraph.
%spin glasses on any graph
%mapping between dimer and spin glasses is only in 2d

In some variations of the dimer model, the edges of the graph carry weights. Such weights can have different interpretations. For example, weights can represent distances been the vertices on a Euclidean domain.%, or weights can be random variables drawn from a distribution.
In such weighted dimer model, the goal is often to find the maximum or minimum weight perfect matching. Finding the maximum weight perfect matching on a graph is a computationally heavy but ``easy'' task (i.e., it belongs to the \textsf{P} computational complexity class) and it is a well studied combinatorial optimisation problem. The problem has also connections with famous (and harder) computational problems, for example the SAT problem and the traveling salesman problem.

%not complete
Moreover, if the weighted dimer model is considered on a two dimensional lattice, its properties have been shown to be in correspondence with the ones of 2-dimensional spin systems (\cite{caracciolo2021criticality}), which are fundamental models in statistical mechanics. Such systems are given on a $2$-dimensional graph, where the vertices are associated to binary variable, called spins, and links join interacting pairs. For example, spins can represent the magnetic charge of a pole, or the direction of voting of the site.

The following review will be structured in three components that outline the key results about the weighted dimer model. First, the so-called infinite temperature limit will be considered. Then some results on the zero temperature limit will be given. Finally, some computational techniques to solve the problem will be reviewed.

\section{Domino tilings} \label{Domino tilings} %infinite temperature

%explain domino tilings from first principles here.

\begin{figure}[h]
    \centering
    \includegraphics[scale=0.1]{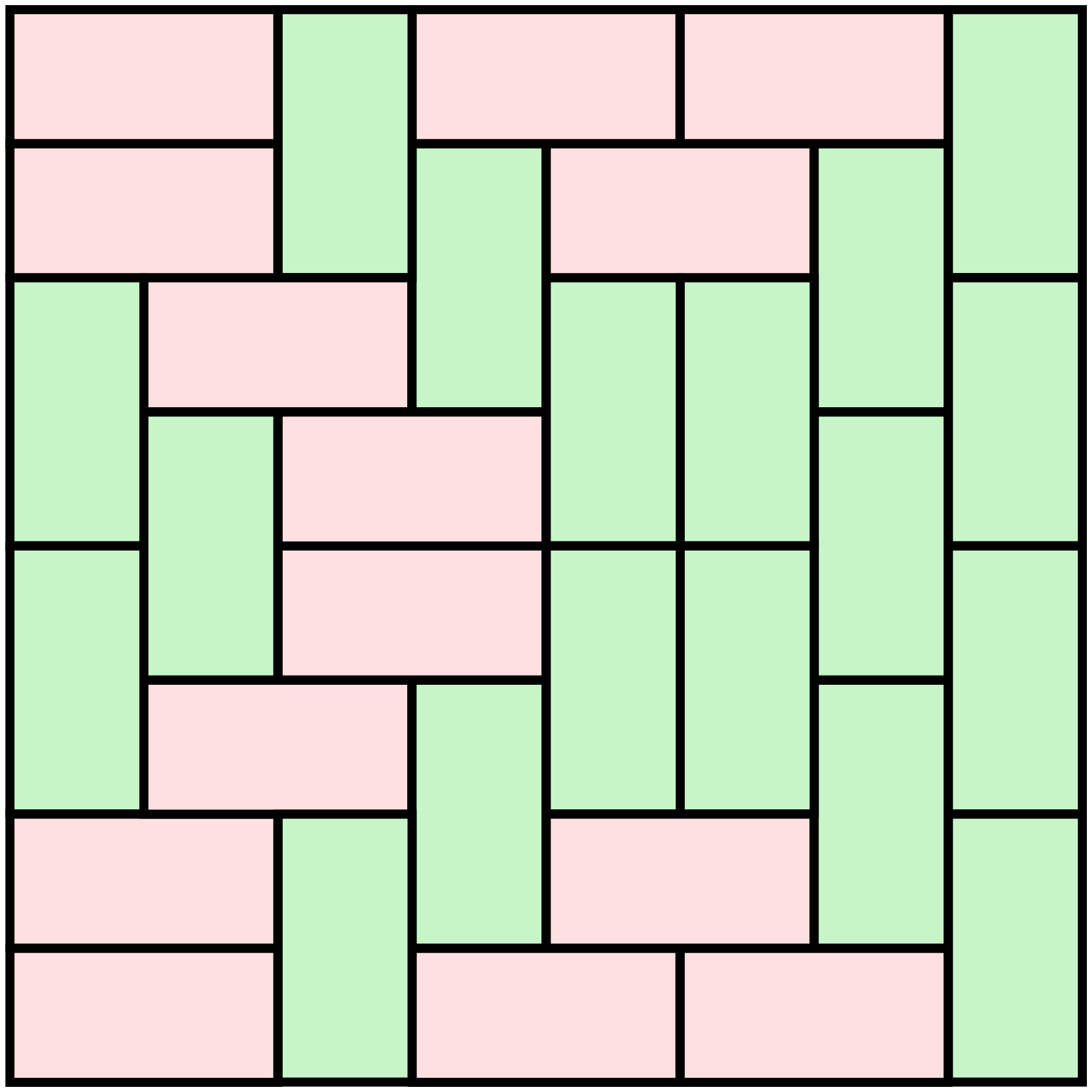}
    \caption{Domino tiling of an $8\times8$ chessboard.}
    \label{fig:1}
\end{figure}

When studied on a subset $G$ of $\mathbb{Z}^2$, the dimer model is equivalent to the tiling problem of a 2-dimensional grid by means of dominos, which are are $2\times 1$ and $1\times 2$ rectangles. A \textit{tiling} is a covering of the grid such that all squares are covered, and there is no overlap between dominos and no dominos hang off of the edges of the grid. See for example Fig.~\ref{fig:1} where $G$ is the $8\times 8$ square grid.

Given $ G\subset\mathbb Z^2$, counting the number of tilings on it is a hard task. One of the most famous results about this problem is the number of domino tilings (perfect matchings) on a finite rectangular lattice, that was obtained independently by Kasteleyn \cite{kasteleyn1961statistics} and Temperley and Fisher \cite{temperley1961dimer} through computing the partition function. 

Let ${G(E,V)}_{m,n}$ be an $ m \times n $ rectangular lattice. The partition function $Z$ of $G$ can be obtained by defining a positive weight $w(e) > 0$ for all edges. Then, the weight of each configuration is:
\begin{equation}
    w(C) = \prod_{e\in C} w(e)
\end{equation}
where the product runs over all edges of a configuration. The partition function is then defined by
\begin{equation}
    Z_G = \sum_C w(C) = \sum_C \prod_{e\in C} w(e)
\end{equation}
where the sum runs over all possible dimer coverings. If we set $w(e) = 1$ for all edges than $Z_G$ is equal to the number of valid dimer coverings on $G$. Kasteleyn showed that the number of dimer coverings of a finite rectangular lattice can be constructed by assigning the following weights to the edges:
\begin{equation}
    K_{u, v} = 
    \begin{cases}
        1,  & \text{if } (u, v) \text{is horizontal edge}\\
        i,  & \text{if } (u, v) \text{is vertical edge}\\
        0,  & \text{otherwise}
    \end{cases}
\end{equation}
from which the number of valid configurations is
\begin{equation}\label{eq:1}
    Z_{m,n} = \sqrt{ \vert \det K \vert} = \left|\prod_{j=1}^m \prod_{k=1}^n 2 \cos \left(\frac{\pi j }{m +1}\right) + 2 i \cos\left(\frac{\pi k}{n+1}\right) \right|^{\frac{1}{2}}.
\end{equation}
%Partition function not defined until now

%where $Z_{m,n}$ is also said to be the partition function of $\mathcal{G}_{m,n}$. The proof of the formula relies on calculating the determinant of the inverse Kasteleyn matrix $K$ of $\mathcal{G}_{m,n}$ and using the identity for the Pfaffian of a matrix $A$: $\mathrm{Pf}(A)^2 =\det(A)$. 
%{\color{red} MAYBE SOME DETAILS ABOUT HOW THE KASTELEYN MATRIX IS CONSTRUCTED, AND WHY THE PFAFFIAN IS IMPORTANT.}
For example, for a chessboard like the one in Fig.~\ref{fig:1}, which is an $8 \times 8$ square lattice, the number of perfect matchings is $Z_{8,8} = 12 988 816$. In the large $m,n$ limit it can be shown that \eqref{eq:1}  simplifies to:
\begin{equation}
    \lim_{m,n \rightarrow \infty} \frac{1}{mn} \log Z_{m,n} = \frac{1}{2 \pi^2} \int_0^{\pi}\int_0^{\pi} 2 \cos \theta + 2 i \cos \phi\,{\mathrm d}\theta {\mathrm d}\phi = \frac{G}{\pi}
\end{equation}
where $G =\frac{1}{1^2} - \frac{1}{3^2} + \frac{1}{5^2}- ... $ is Catalan's constant.

\subsection{Height function and conformal invariance}
\begin{figure}[h] 
  \centering
  \begin{minipage}[b]{0.4\textwidth}
    \includegraphics[width=\textwidth]{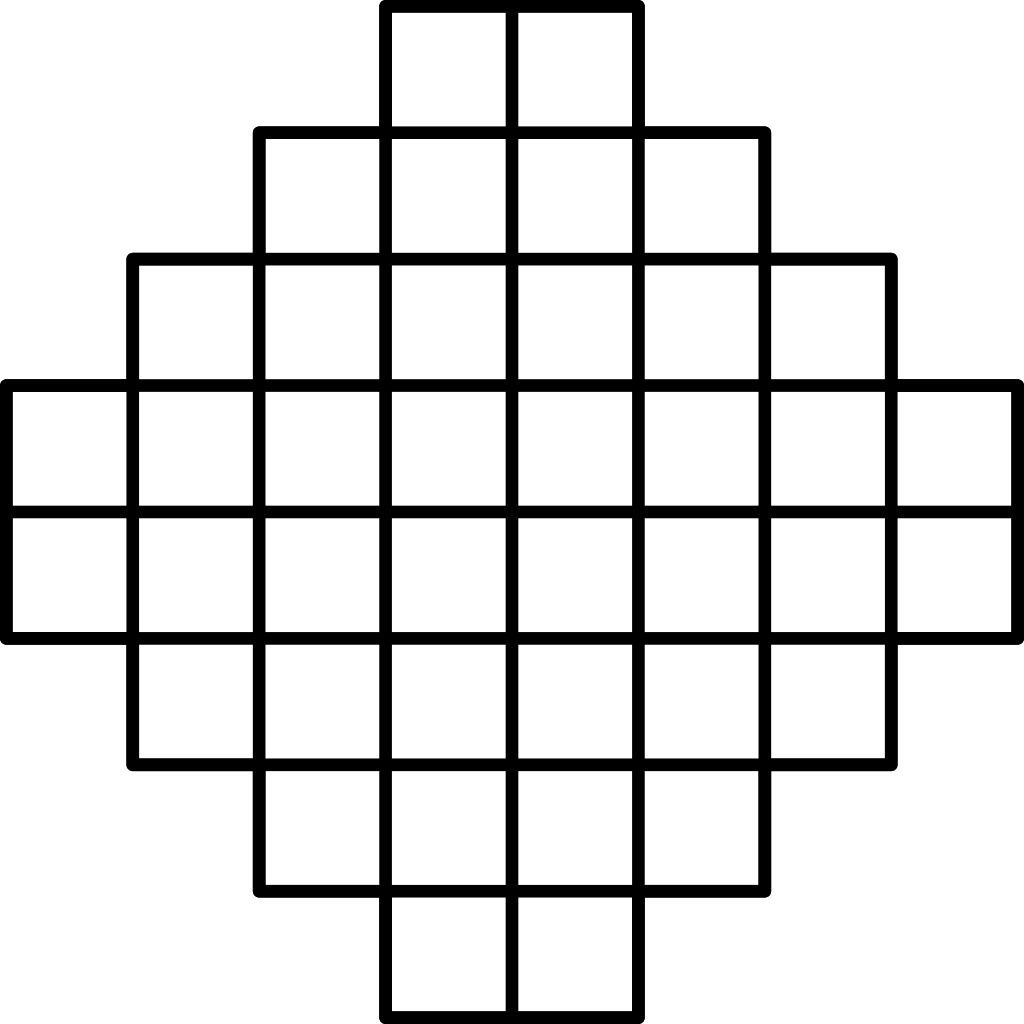}
  \end{minipage}
  \hspace{25pt}
  \begin{minipage}[b]{0.4\textwidth}
    \includegraphics[width=\textwidth]{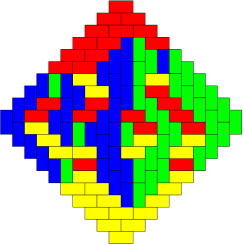}
  \end{minipage}
  \caption{The Aztec diamond domain with an example of a covering that shows the arctic circle phenomena.\label{Aztec}}
\end{figure}

%include figure
%explain uniform more precisely
Domino tilings on a rectangular grid are homogeneous. Roughly speaking, this means that a randomly drawn tiling will look random from the set of all tilings to the human eye. However, this is not the case for all graph structures. Consider the Aztec Diamond shape, shown in Fig. \ref{Aztec}. Given a tiling uniformly and randomly extracted from the set of all possible tilings, a frozen region will likely appear, where the dominos around the edges align, thus the covering is no longer uniform. This is the so-called arctic circle phenomenon, and it has been studied by means of an auxiliary tool, the height function \cite{kenyon2000conformal}.

The height function is a $\mathbb{Z}$-valued function defined on the vertices of the tiling, and it is given up to a global additive constant. The height of a domino covering of a graph that has a chessboard-like coloring of black and white squares can be calculated with an iterative procedure: start with any vertex and set the height there to 0. Then by following the edges of the domino tiling calculate the height of subsequent vertices. Increase the height by $+1$ if there is a black square to the left of the edge, and decrease the height by $-1$ if there is a white square to the left of the edge. If there is no square to the left of the edge, then do the opposite for the color of the square on the right. The height function of a tiling is then a map from $\mathbb{Z}^2$ to $\mathbb{Z}$. A domino tiling uniquely defines a height function and vice versa. The coordinates given by the height function outline a surface with support on the surface that has been tiled. The artic circle phenomenon is then studied in terms of the statistical properties of the set of feasible height functions on the given domain. 

In \cite{kenyon2000conformal} the authors show that, in the limit of lattice spacing going to $0$, the height distribution of a random tiling is \textit{conformally invariant}, and thus certain macroscopic properties of the tiling itself are conformally invariant. Conformal invariance is strong form of $2$-dimensional invariance that emerges in statistical mechanics models.  It can be broken down into three weaker forms of symmetries: translational symmetry, rotational symmetry and scale invariance. All three invariance properties need to be satisfied for conformal invariance. For systems with disorder, conformal invariance may appear after averaging over multiple instances of the system, which removes the instance specific heterogeneity.

Proving conformal invariance in probabilistic and physical models has been of central importance for mathematicians and physicists. The presence of conformal invariance allows for the application of a large set of tools of conformal field theory and the reduction of the model themselves to very few informative numbers corresponding to the so-called universality classes of the models themselves. In the 1970s Alexander Polyakov predicted that lattice models at criticality exhibit this invariance but his statement has remained unproven for decades. S. Smirnov was able to prove conformal invariance in the scaling limit for percolation on the triangular lattice in \cite{smirnov2001critical} and later for the Ising model at criticality \cite{smirnov2007conformal}. For his work on conformality of lattice models he received the Fields medal. Conformal invariance for the loop-erased random walk and the uniform spanning tree, and their convergence towards a special family of conformally invariant processes, called SLE processes (see below), was shown by O. Schramm in \cite{schramm2000scaling}.

With respect to the dimer model, R. Kenyon showed that the measure of the height function of the dimer coverings is conformally invariant as the lattice spacing tends to zero. Furthermore, in \cite{kenyon2014conformal} Kenyon proves that the loops in the double dimer model are conformally invariant and converge to an $\mathrm{SLE}_4$ process.

\subsection{Schramm-Loewner Evolution}
The Schramm-Loewner evolution (SLE), indexed by parameter $\kappa$ as $\mathrm{SLE}_\kappa$,is family of stochastic processes in the complex plane. There are two main variants: chordal SLE, which give a family of curves between two fixed boundary points on the upper half of the complex plane $\mathbb{H}$ and radial SLE which gives a family of curves between a fixed boundary and fixed interior point on the unit disk $\mathbb{D}$.

SLE was discovered by Schramm in \cite{schramm2000scaling} and later developed by W. Werner and G. Lawler who were awarded the Fields medal for their work.
%could include exact descriptions
Models such as the loop-erased self avoiding random walk, the uniform spanning tree, the critical Ising model, the dimer model, the double dimer model and percolation Potts model have been shown to belong to $\text{SLE}_{\kappa}$. An $\text{SLE}_\kappa$ curve $\gamma(t)$ defined on the upper half complex plane $\mathbb{H}$ is given by $\gamma(t) = g_t^{-1}(\xi_t)$ where $g_t$ solves the Loewner differential equation
\begin{equation}
    \frac{\partial g_t(z)}{\partial t} = \frac{2}{g_t(z) - \xi (t)}
\end{equation}
with boundary condition $g_0(z) = z$. The equation depends on the driving function $\xi(t)$, which is a Brownian process with mean $\langle \xi(t) \rangle = 0$ and variance $\langle \xi^2(t)\rangle = \kappa t$. The curves satisfy conformal invariance: they are self-similar under conformal mappings. The curves also satisfy domain Markov property: given a curve $\gamma_{ab}$ that lies between two boundary points $a$ and $b$, the domain Markov property states that the probability of a segment $P_D(\gamma_{cb})$ conditioned on the rest of the curve $\gamma_{ac}$ is identical to the probability of the original curve on the cut domain $D \setminus \gamma_{ac}$ conditioned on starting at $c$
\begin{equation}
    P_D (\gamma_{cb} \vert \gamma_{ac}) = P_{D\setminus \gamma_{ac}} (\gamma_{cb} \vert c).
\end{equation}
%{\color{red} THE DOMAIN MARKOV PROPERTY DESCRIBED HERE WAS WRONG. SUBSTITUTE WITH A CORRECT STATEMENT...}

\begin{figure}[h]
    \centering
    \includegraphics[width=0.95\textwidth]{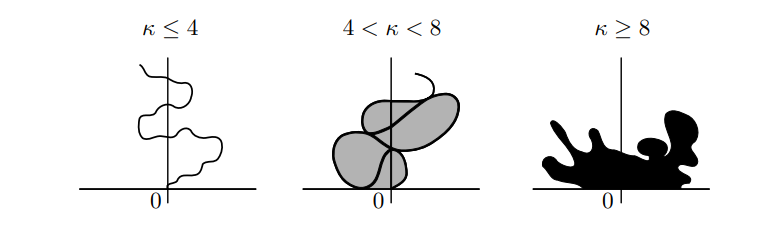}/
    \caption{SLE curves for different values of $\kappa$. Image taken from \cite{fabbricatore2021critical} }
    \label{fig:SLE}
\end{figure}

From the definition it is clear that the only parameter of the process is $\kappa$. It can be shown that for $\kappa \le 4$ the curves are simple with probability 1, for $4 < \kappa < 8$ the curves have double points and all points are contained within a loop and for $\kappa \ge 8$ the curves are space filling with probability 1, see Fig.~\ref{fig:SLE}. Finally, SLE objects are fractal objects. The formula that connects the variance of the driving function $\kappa$ and the fractal dimension of the curve $d_f$ is $d_f = 1 + \frac{\kappa}{8}$.

\section{Criticality and conformality in spin glasses}

%talk about connection to dimer model and ising model

There exists a one-to-one correspondence between the 2-dimensional Ising model and the 2-dimensional dimer model. This bijective mapping was developed by Fisher in \cite{fisher1966dimer}. The method maps the Ising model on a graph $ G$ embedded in the torus to the dimer model defined on a decorated version of $ G^F$ of $ G$, called \textit{Fisher graph}. By means of this mapping, the study of the Ising model reduces to the study of the dimer model on the Fisher graph, which explains why the dimer model has become of such high importance amongst physicists. For a full description and proof of the mapping see \cite{fisher1966dimer}. This fact also motivates the following short overview on spin glasses, and a discussion of the characteristics of such models at the critical temperature.
%could cite here something newer.

%talk about criticality and universality of models. Why they are important and where we can find, common signs of them.

\subsection{The Ising model and spin-glasses}

%at a temperature different from zero
Spin glasses are disordered magnetic materials characterized by randomness in the interactions. Before giving a more precise definition of a spin-glass, let us introduce the standard Ising model, a cornerstone of statistical mechanics. The Ising model is defined on a graph $ G$, where each vertex $i$ is associated to a spin $\sigma_i \in \{\pm 1\}$. A spin configuration $\mathbf{\sigma}$ is an assignment of spin to each lattice site. The energy of a spin configuration is given by the Hamiltonian:
\begin{equation}
     H = - \sum_{\langle i,j \rangle} J_{ij} \sigma_i\sigma_j - \sum_j h_j \sigma_j
\end{equation}
%specify what J and h are
where the sum $\langle i,j \rangle$ runs over nearest neighbour spins. The configuration probability is given by the Boltzmann distribution for the fictitious inverse temperature parameter $\beta = \frac{1}{T}$
\begin{equation}\label{eq:Boltz}
    P_\beta(\sigma) = \frac{1}{Z} e^{-\beta H(\sigma)}
\end{equation}
where $Z$ is the normalisation constant, or partition function:
\begin{equation}
    Z = \sum_\sigma e^{-\beta H(\sigma)}.
\end{equation}
L. Onsager provided a solution to the partition function of the Ising model on a two dimensional square lattice in 1944 with the celebrated transfer matrix method assuming $J_{ij}\equiv J$. The model has a phase transition at finite temperature $T_c=\frac{2J}{\ln(1+\sqrt 2)}$.

%sentence sounds strange writes
Ising spin glasses are Ising spin systems with quenched disorder in the couplings $J_{ij}$. A system is quenched when the parameters describing disorder do not depend on time, they are "frozen" in time. A typical example of a quenched spin glass is the Edwards-Anderson model with the Hamiltonian
\begin{equation}
    H = - \sum_{\langle i,j \rangle} J_{ij} \sigma_i\sigma_j
\end{equation}
defined on a $d$-dimensional hypercubic lattices, where $J_{ij}$ are Gaussian random variables. Disorder creates frustration in systems as it becomes impossible to satisfy all of the couplings. That means that if from initial spin we fix the other spins in a chain in the graph $ G$, opposing effects and thus a contradiction might arise. To deal with disorder, self-averaging quantities are studied. Self-averaging quantities do not depend on the specific realisations of the system when the size $N$ of the system goes to infinity, and allow the building of the theory of the spin glasses. For example the free energy is a self-averaging quantity
\begin{equation}
    F\coloneqq \lim_{N \rightarrow \infty} F_N(\beta, J) = \lim_{N \rightarrow \infty} -\frac{1}{\beta N} \log \int d\sigma e^{-\beta H(\sigma, J} = F_\infty(\beta)
\end{equation}
where in the $N \rightarrow \infty$ limit $F$ no longer depends on the random realisation of $J$ (but it does on the distribution of $J$). The computation of $F$ usually requires the use of sophisticated techniques. The result is expressed in terms of the distribution of order parameters taking the form of \textit{overlaps}. The overlap quantifies the similarity between two configurations of spins. Given two configurations $\sigma$ and $\tau$ the overlap between them is
\begin{align}
    q_{\sigma \tau} = \frac{1}{N} \sum_{i=1}^N \sigma_i \tau_i
\end{align}
which is $1$, if the two configurations are identical, $-1$ if completely opposite and $0$ if they are uncorrelated for Ising spins. 

%Change title 
\subsection{Phase transitions}

A phase transition is a qualitative change in the state of a system under a continuous change in an external parameter. A physical example of a phase transition is when water reaches the boiling point it evaporates, resulting in an abrupt change in its volume, thus going through a phase transition. The point where the phase transition occurs is called the critical point. At the critical point the two phases of the material have equal free energies thus are equally likely to exist. There are two main classes of phase transitions: first order phase transitions that involve latent heat; and second order phase transitions or continuous phase transitions, with no latent heat but diverging correlation length at criticality.

The aforementioned EA model exhibits a phase transition at a critical temperature $T_c$ in the $N \rightarrow \infty $ limit. In this limit, the length scale governing fluctuations -- the correlation length $\xi$ -- diverges to infinity. At this critical point, observables such as the heat capacity and susceptibility show singularities. As a consequence of diverging correlation lengths, the macroscopic properties of the model become highly nonlinear.
%as the interactions encoded in the Hamiltonian $H$ are affected by spins globally. 
%{\color{red} WHAT DOES THIS SENTENCE MEAN?}

To better understand the properties of the EA in $2d$ at criticality, Hartmann and Young \cite{hartmann2001lower} investigated the EA model. Consider the Hamiltonian
\begin{equation}\label{eq:2}
    \mathcal{H} = - \sum_{\langle i,j \rangle} J_{ij} \sigma_i \sigma_j
\end{equation}
where the sum runs over all pairs of nearest neighbours on the 2D square lattice, $\sigma_{i} = \pm 1$ and $J_{ij}$ are quenched random variables. Two kinds of disorder are considered:
\begin{enumerate}
    \item $J_{ij}$ is a standard normal random variable;
    \item $J_{ij} = \pm 1$ with equal probabilities (bimodal distribution).
\end{enumerate}

By means of a polynomial time algorithm to calculate the minimum-weight perfect matching, and using the mapping between $2d$ Ising model and matching, they were able to efficiently compute the ground state energy $E_0$ of the system.
%which has energy $E_0$ that is equal to the sum of weights in the perfect matching $E_0$.
%last sentence is dodgy above
At this point, one of two types of perturbations is introduced in the model: (i) changing the periodic boundary conditions to anti-periodic, which means swapping the signs of the edges that cross the boundary and (ii) changing the free boundary conditions to periodic ones. %{\color{red} WHICH BOUNDARY!?!?} 

The excitation forces the system to move from its former lowest energy ground state to a higher energy excited state. Note, that in context of lattice models excitations can take many forms, for example the removal of links/vertices, changing the weights on the edges or changing the boundary conditions are all different forms of excitations. After the perturbation the new perturbed energy $E_0^{\text{pertrb}}$ of the model is calculated by rerunning the algorithm. The average difference in the energy is given by $\Delta E = \langle E_0^{\text{pertrb}} - E_0 \rangle$. It is expected that the change in average energy will vary as a function of the system size $L$ as $\Delta E \sim L^\theta$. Here $L$ is the number of vertices in one direction and $\theta$ is referred to as the stiffness exponent. The stiffness exponent is informative about the criticality properties of the system: indeed, the correlation length $\xi$ is expected to diverge at zero temperature for $\theta \le 0$. The exponent $\theta$ was estimated in both settings. For the Gaussian disorder the numerical results match previous results that $\theta < 0 $, hence the lower critical dimension, which is the largest dimension at and below which phase transitions do not occur is $d_c > 2$. 
%{\color{red} EXPLAIN WHAT IS THE LOWER CRITICAL DIMENSION}.
Surprisingly, the $\Delta E$ in the Ising model with Bimodal disorder is found to saturate at a non-zero value for $L \rightarrow \infty$ which indicates that $\theta = 0$, hence the lower critical dimension is exactly $d_c = 2$.  Criticality and then further universality\footnote{Universality says that critical exponents do not depend microscopic properties but only on broad features, for example the dimensions of the problem or the type of randomness in the model.} in the 3-dimensional Edwards-Anderson model and the Ising model with bimodal disorder has been also investigated in a large scale Monte Carlo simulation in \cite{Katzgraber_2006}. All models are shown to belong to the anticipated universality class by estimating the overlap between the models and their excited counterparts. Strong finite size corrections are present in the study.
%{\color{red} SO? WHAT CONCLUSIONS?}

\subsubsection{SLE in spin glass models}

%SLE should be defined until here
%Mention domain markov property
Proving the relation between SLE and Ising spin glasses at criticality has attracted large efforts over the years. If true, it allows the usage of methods from stochastic calculus and conformal field theory, which are rich areas of research themselves. SLE curves need to satisfy two properties: conformal invariance and domain Markov property.

%disorder is not explained in the model
%rewrite this in passive.
In \cite{amoruso2006conformal} and \cite{bernard2007possible} various numerical tests are applied to study the conformal invariance of domain walls in the 2-dimensional Ising spin glasses with Gaussian bond distribution at criticality. %The domain wall is interface separating different magnetic domains.
Even though there is random disorder in the model on the microscopic level, the divergent correlation length suggests that at the critical point long range interactions are dominant and conformal invariance could arise. 

In particular, in \cite{amoruso2006conformal} a conformality test is applied to the domain wall based on the hypothesis that the domain wall will cross a predetermined fixed line in the domain $n$ times with the same probability as it crosses that same line $n$ times after the application of a series of conformal maps to the original geometry. The naive expectation would be that because the conformal transformation of the geometry introduces slits (on purpose) in the transformed geometry, the probability of crossings decreases. In actuality, the opposite is found. As system size increases, the probability of domain wall crossings show convergence between the two conformally equivalent geometries. The domain wall is mapped then to the complex upper half plane. Assuming then that the domain wall is indeed a SLE process, the variance of the driving function $\kappa$ is estimated to be $\kappa \approx 2.1$. % which is consistent with findings in \cite{fabbricatore2021critical}. 
Furthermore, a relationship between the stiffness exponent $\theta$, which was calculated in \cite{hartmann2001lower} and the fractal dimension $d_f$ of the domain wall is found to satisfy the relation:
\begin{equation}
    d_f = 1 + \frac{3}{4(3+\theta)}
\end{equation}

In \cite{bernard2007possible} several different SLE tests are applied to domain walls of the EA model on a triangular lattice on the annulus. The fractal dimension $d_f$ is estimated from a scaling ansatz and is found $d_f = 1.28$, which is consistent with previous literature. For the conformal invariance of the domain wall, an important result from the theory of SLE curves is used: without going into too much detail, it is possible to calculate the probability that an SLE curves passe a point in the complex plane to the left \cite{schramm2001percolation}. This probability is calculated for the domain walls, and matches the expectations. The diffusion parameter $\kappa \approx 2.32$ is slightly higher than in \cite{amoruso2006conformal}, but remains within the range commonly quoted in other papers. Finally, a test for the Markovian property of the driving function $\xi(t)$ was performed. The driving function $\xi(t)$ was obtained by mapping the domain wall to a real valued sequence $\xi(t_i)$ and then interpolated. For SLE, the generated curve is a Brownian motion with diffusion parameter $\kappa$.
%{\color{red} NOT CLEAR: HOW $\xi$ WAS OBTAINED?}: 
The correlation between subsequent values of $\xi(t)$ defined as: $C_d(n) = \langle (\xi(t_{i+n+1}) - \xi(t_{i+n}) (\xi(t_{i+1})- \xi(t_{i} ) \rangle $ is calculated; and found to decay rapidly for $n=2$ to $n=8$. Note, that due to the structure of the problem, some edges are forbidden which induces some short term correlation, hence the driving function is Markovian. The test for the domain Markov property of the system after averaging fails asymptotically, however only small system sizes were checked thus for larger systems the property might still hold.

%\subsubsection{Other estimates  the fractal dimension}

%

Excitations or perturbations of the EA model have been also used to estimate the fractal dimension of the domain walls (studying the induced rearrangements) or their robustness.

In \cite{Corberi_2019}, for example, a specific quenched Ising model is considered, namely the bond diluted Ising model (BDIM). This is a ferromagnetic spin system on the square lattice where pairwise interactions between vertices are nullified with probability $d$. The Hamiltonian of the system is as \eqref{eq:2}, where $J_{i,j}$ are i.i.d.~random variables with $P(J_{i,j}) = (1-\rho)\delta_{J_{i,j}, J} + \rho \delta_{J_{i,j}, 0}$. The aim of the paper is to study the interface between opposite magnetisations, and its dimension $d_f$, as a function of the dilution parameter $\rho$. The ``clean state'' corresponding to $\rho=0$ is used as a reference point. The quench is introduced at time $t=0$ in the infinite temperature equilibrium state and then evolved with Glauber update rates:
\begin{equation}
    w(\sigma_i \rightarrow -\sigma_i) = \frac{1}{2} \left[1 - \sigma_i \tanh{\frac{\sum_{\langle i,j\rangle}J_{ij}\sigma_j}{T}}   \right]
\end{equation}
where $T$ is the final temperature of the quench. The average size of the growing domain $R(t,\rho)$ for $\rho=0$ satisfies the scaling law 
$$R(t,0) \sim t^{\frac{1}{2}}$$
however, for $ 0 < \rho < \frac{1}{2} = \rho_c$ the curve is dependent on $\rho$, and the growth is slower compared to the clean state. At this point, the fractal dimension is extracted by measuring the average squared winding angle $\langle \theta^2 (r,t,\rho) \rangle $ between points at distance $r$ on the perimeter of the domain wall. The winding angle variance is known to behave as $\langle\theta^2(r)\rangle\sim 4\left(1-\frac{1}{d_f}\right)\ln r$ for self-similar interfaces (and in particular, as we will discuss, SLE curves). The authors show that at dilution $\rho = 0$ the fractal dimension of the domain wall is $d_f = \frac{4}{3}$. With dilution $\rho>0$ the relationship between the distance $r$ and $\langle \theta^2 (r,t,\rho) \rangle $ can be deconstructed in two different regimes depending on the ratio $r/r_{\text{cross}}$, where $r_{\text{cross}}$ is a crossover scale. This is explained by the the natural length of the domain $R(t,\rho)^{d_f}$ such that  $\langle \theta^2 (r,t,\rho) \rangle $ behaves differently for  $ r\ll R(t,\rho)^{d_f}$ and for $r \gg R(t,\rho)^{d_f}$. Consequently, the fractal property of the domain is different on short vs. long scales and is dependent on dilution $\rho$. This result is at variance with the fractal dimension of the random bond Ising model, where the fractal dimension is described by a unique exponent irrespective of the bond disorder. Even though the BDIM is a similar model to the random field and random bond Ising model, they do not belong to the same universality class. The BDIM shows similar scaling on measured observables: crossing probabilities, averaged squared winding angle and pair connectedness, however the scaling is always dependent on the disorder, which is not the case for the reference models. In the following we will show more precisely how the fractal dimension of general SLE can be extracted by studying the behavior of the variance of the winding angle $\langle\theta^2\rangle$.

%{\color{red} THE REASON OF THIS SECTION IS NOT CLEAR}
\subsubsection{SLE in the FK random cluster model}
The winding angle variance is a good tool for the study of fractal dimensions of SLE curves. For example it has been considered in \cite{wieland2003winding} for the contours in the 2D Fortuin-Kasteleyn (FK) random cluster model. The FK cluster model is a bond percolation model with bond dilution parameter $p$ with an additional parameter $q$ on clusters. The variance of winding angle at points where $k$ strands of the perimeter of the cluster connect as well as for the external perimeter. The winding function $r(\bullet)$ on the edges is constructed by starting from an arbitrary edge $e$ on the perimeter and then setting the value of $r(e')$ at a neighbouring edge to $r(e') = r(e) + \text{turning angle between }(e,e')$. For this model too, the connection
\begin{equation}
    d_f = 1 + \frac{\kappa}{8}
\end{equation}
between the wining angle variance, the fractal dimension and $\text{SLE}_\kappa$ is confirmed to hold. 

\begin{figure}[h]
    \centering
    \includegraphics[width=0.8\textwidth]{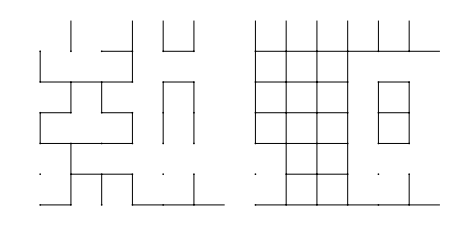}
    \caption{Contours of the perimeter and external perimeter of the FK cluster model. Image from \cite{wieland2003winding}}
    \label{fig: FK Cluster}
\end{figure}

\subsubsection{SLE in the random dimer model}

Until this point in the chapter, we have seen that curves arising in various lattice models exhibit conformal invariance as well as similarities in their fractal dimension are found. Utilising the techniques and ideas from those papers, the authors of \cite{caracciolo2021criticality} take into consideration the random dimer model. For a formal introduction to the model see Ch. \ref{chap: max weight}. 

Random link excitations are applied to the random dimer model, such that a new optimal configuration arises. Once overlaying the ground state configuration and the optimal excited configuration, a cycle appears on the lattice. It is shown that the ground state solution of the RDM is at criticality and furthermore, the cycles show signs of conformal invariance. There is evidence in \cite{caracciolo2021criticality} that the RDM on a bipartite lattice - square grid, hexagonal grid and random euclidean assignment problem - belong to the same universality class as the 2-dimensional self-avoiding random walk with  $D_f \simeq 1.252$ and are compatible with an $\mathrm{SLE}_\kappa$ with $\kappa = 2$, while the monopartite models - triangular lattice and random euclidean matching problem - are compatible with domain walls in 2d spin glasses with $D_f \simeq 1.274$.

Finding the ground state of the random dimer model is analogous to a combinatorial optimisation problem, namely the minimum weight perfect matching. In the next section Sec. \ref{Optimisation} a short overview of an algorithm that finds the optimal configuration from an optimisation perspective is presented.

\section{Optimisation toolbox} \label{Optimisation}

In the weighted dimer model the objective is no longer to count all the possible dimer coverings, but to find the maximum (or minimum, the problems are equivalent) weight perfect matching. The maximum weight perfect matching is defined as the covering that maximises the sum of weights that are in the covering. However, finding the maximum weight perfect matching is infeasible by first finding all possible domino tilings, and then selecting the one with the maximum weight as the number of configurations rise rapidly with system size. Combinatorial optimisation techniques have been developed to solve this problem.

Combinatorial optimisation theory is a subfield of mathematics that deals with searching for the maxima of an objective function whose domain is a discrete but large configuration space. Famous examples in the field are the traveling salesman problem, where the objective is, given a list of cities and distances between all of them, what is shortest path that connects $N$ cities exactly once; or the boolean satisfiability problem, where the objective is to assign values to a set of variables to satisfy a given boolean expression.

Solving the assignment problem or finding the maximum weight perfect matching of a graph is also a well studied combinatorial optimisation problem. As shown in \cite{kasteleyn1961statistics}, the number of domino tilings on a grid is very large, thus the brute force approach of looking at all the possible combinations to find the one with the lowest sum of weights is not feasible. Instead, polynomial time algorithms have been developed that solve the assignment problem. That makes the problem \textsf{P} hard\footnote{The \textsf{P} vs. \textsf{NP} problem is one of the unsolved Millennium Prize problems that asks whether every problem whose solution can be quickly verified can also be solved quickly.}. Edmond's blossom algorithm is one example of a polynomial time algorithm that solves the assignment and matching problems. This is the algorithm adopted in this analysis presented in this report, and it is briefly discussed below.

\subsection{Edmond's blossom algorithm} \label{Edmonds}

Edmond's blossom algorithm was developed by Jack Edmonds in 1961 and published in 1965 in \cite{edmonds1965i965a}. The algorithm finds the maximal matching of a graph $G = (V, E)$ in time complexity $O(V^2E)$. Since the original paper, multiple improvements have been added to improve performance, for example \cite{kolmogorov2009blossom}. Here, only the unweighted case is described for brevity, see \cite{kolmogorov2009blossom} for a full description.

In the unweighted case, the algorithm starts with a random matching $M$ and then iteratively finds edges to improve by scanning for an augmenting path $P$. An augmenting path is an alternating path that starts and ends on exposed vertices and alternates between edges that belong to the matched set $e \in M$ and edges that do not $e\notin M$. A vertex $v$ is exposed if no edge $e\in M$ is incident to $v$. The algorithm takes advantage of Berge's lemma, which says that a matching $M$ is maximal iff there is no augmenting path in $M$. If an augmenting path is found a new matching is modified to $M_{\text{new}}= M \otimes P$. The algorithm is then iterated until no augmenting path is found, at which point there are 2 possibilities: (i) $M$ is not a possible maximal matching and no maximal matching exists thus the algorithms exits or (ii) a maximal matching is found and the algorithm returns $M$.

A central part of the algorithm are blossoms, which is where the name comes from.  Blossoms reduce cycles of odd length to a single vertex so that the search for augmenting paths happen on the contracted graph.
\begin{figure}[h]
    \centering
    \includegraphics[scale=0.5]{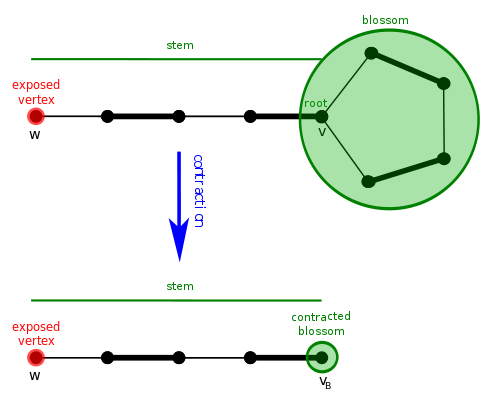}
    \caption{Illustration of how the blossoms are contracted into a single vertex. Image from \protect\url{https://brilliant.org/wiki/blossom-algorithm/}}
    \label{fig:my_label}.
\end{figure}

Blossoms improve the efficiency of finding augmenting paths, as exposed vertices are guaranteed not to be in a blossom. Blossoms also allow the Hungarian algorithm, discussed below, to be deployed, which is defined on bipartite graphs only.

\subsection{Hungarian algorithm}
%extend this too dense
The Hungarian matching algorithm finds the maximum weight perfect matching on a bipartite graph. One of its limitations is that it is constrained to bipartite graphs, however the blossom algorithm generalises that to monopartite graphs as well. It was named after Hungarian mathematicians Dénes Kőnig and Jenő Egerváry for their work on graph theory that forms the basis of the algorithm.

\textbf{Definitions:}
\begin{itemize}
    \item A \textit{labeling} on a graph $G=(V,E)$ is a function $l\colon {V} \rightarrow \mathbb{R}$ such that
    \begin{equation}
        \forall (u,v) \in E: l(u) + l(v) \ge \text{weight}(u,v)
    \end{equation}
    \item Kuhn--Munkres Theorem: Given a labeling $l$, if $M$ is a perfect matching on $G_l$, than it is the maximum-weight perfect matching. 
\end{itemize}

The algorithm works by iteratively improving the labeling on the graph to take advantage of the Kuhn--Munkres theorem. Let $M'$ be any perfect matching on $G$. Then, by the definition of the labeling function
\begin{equation}
    \text{weight}(M') =  \sum_{(u,v)\in M'} \text{weight}(u,v)
    \le \sum_{(u,v)\in M'} l(u) + l(v) =  \sum_{v \in V} l(v)
\end{equation}
This means that $\sum_{v \in V} l(v)$ is an upper bound on the cost of any perfect matching. Thus, the maximum weight perfect matching $M$ will satisfy
\begin{equation}
    \text{weight}(M) = \sum_{(u,v)\in M} \text{weight}(u,v) =\sum_{v \in V} l(v) 
\end{equation}

Hence, the algorithm works as follows: start with a matching $M$ and a valid labeling $l$. Look for an augmenting path in $M$. If there is no augmenting path and $M$ is not the maximum weight matching, than improve the labeling $l$ and then repeat from the previous step. The algorithm converges deterministically and has time complexity $O(V^3)$.

\chapter[Maximum-weight excitations in random dimer coverings]{Maximum-weight excitations\\in random dimer coverings} \label{chap: max weight}

\section{Random Dimer Model} \label{RDM}
The random dimer model is the disordered version of the  dimer model. Consider the dimer model on a graph $G(V,E)$ where $V$ is the vertex set and $E$ is the edge set, and assign weights to all edges $e\in E$ from a continuous distribution $q(w)$ as independent and identically distributed random variables. A dimer covering of the model is a subset of edges such that every vertex is covered exactly once. We can assign a cost $E[D]$ for all dimer coverings of the graph as
\begin{equation} \label{eq:3}
    E[D] = \sum_{e\in D} w_e.
\end{equation}
The minimum weight perfect matching corresponds to the covering that minimises \eqref{eq:3} and is the ground state of the random dimer model. From a thermodynamic point of view, the partition function of the random dimer model can be formulated as 
\begin{equation}
    Z(\beta) = \sum_{D} e^{-\beta E[D]}
\end{equation}
for a fictitious inverse temperature parameter $\beta = \frac{1}{T}$. The $Z(0)$ configuration counts the number of dimer coverings which was discussed in further detail in the  Sec. \ref{Domino tilings}. The ground state configuration of the random dimer model is defined as:
\begin{equation}
    E[D^*] = \min_D E[D] = - \lim_{\beta \rightarrow \infty} \frac{1}{\beta} \ln Z(\beta).
\end{equation}

\begin{figure}[h]
    \centering
    \includegraphics[width = 0.8\textwidth]{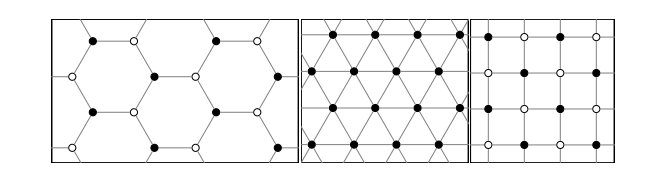}
    \caption{Examples of the honeycomb (H), square (Q) and triangular (T) lattices for $L=4$. Image from \cite{caracciolo2021criticality}.}
    \label{fig:lattice types}
\end{figure}

In this paper lattice graphs are examined, where multiple viable dimer coverings exist. Given that weights are randomised from a continuous probability distribution $q(w)$, the ground state solution of the model is almost surely unique. Three different kinds of lattices are considered. The honeycomb (H) lattice, the triangular (T) lattice and the square (Q) lattice, see Fig. \ref{fig:lattice types}. Periodic boundary conditions are imposed in all directions, modifying the graph topology from a square into a torus $T^2$, which is a product of 2 circles $T^2 = S^1 \times S^1$. 

Each lattice has $L \times L $ number of vertices placed down in the desired shape such that all edges have length 1. Note that the H and Q lattices are bipartite and thus $2N = L^2$ vertices are generated, whereas the T lattice is monopartite and $N=L^2$ vertices are generated.

Finding the ground state of the random dimer model by finding all possible  dimer coverings and then selecting the one with the lowest cost is infeasible due the vast number of configurations. Instead, Edmonds' Blossom algorithm is used that finds the maximum weight perfect matching by iteratively selecting the optimal edges. For an overview of how the algorithm works see Sec. \ref{Edmonds} on combinatorial optimisation. The implementation of the blossom algorithm is executed by the LEMON C++ library, created by \cite{dezsHo2011lemon} that has $O(E V log(V)$ time complexity, where $E$ and $V$ are the number of edges and vertices of the graph.
The code ran on a machine with 4 cores and 8GB RAM memory.

\section{Maximum weight excitation}

In this chapter, we introduce the random dimer covering problem using a `maximum weight excitation'. The maximum weight excitation that we will investigate is a variation of the random link excitation studied in \cite{caracciolo2021criticality}. As in the model therein, each edge weight $w_e$ is randomly drawn from an exponential distribution $q(w) = e^{-w}$, which completely defines the model setup. The ground state solution $D^*$ is found numerically solving for the optimal dimer covering. Subsequently, the edge with the maximum weight $$\hat{e} = \arg\max_{e\in D^*} w_e $$
is deleted. Observe that in \cite{caracciolo2021criticality} a random edge is deleted. The matching algorithm is then rerun on the lattice where $\hat e$ has been removed to find the new optimal solution $D_{\hat{e}}^*$ which therefore will not contain the edge $\hat e$. The new optimal solution is necessarily at a higher cost than the ground state, as otherwise the ground state would be incorrect. The difference is defined as:
\begin{equation}
    \Delta E_{\hat{e}} = E[D_{\hat{e}}^*] - E[D^*]
\end{equation}
which is always non-negative. To analyse the effect of the perturbation, we consider the symmetric difference between $D_{\hat{e}}^*$ and $D^*$:
\begin{equation} \label{eq:sym diff}
    S_{\hat{e}} = \{e\in E : e \in D_{\hat{e}}^*\triangle D^* \}
\end{equation}
where the symmetric difference operator selects the edges that are either in the ground state or the excited state, but not in both. The procedure is repeated for a number of iterations: assigning a set of random weights to taking the symmetric difference between the excited and ground state defines one iteration of the model. The output of such an iteration is a closed cycle $S_{\hat{e}}$. 

\begin{figure}[h]
    \centering
    \includegraphics[width = 0.7\textwidth]{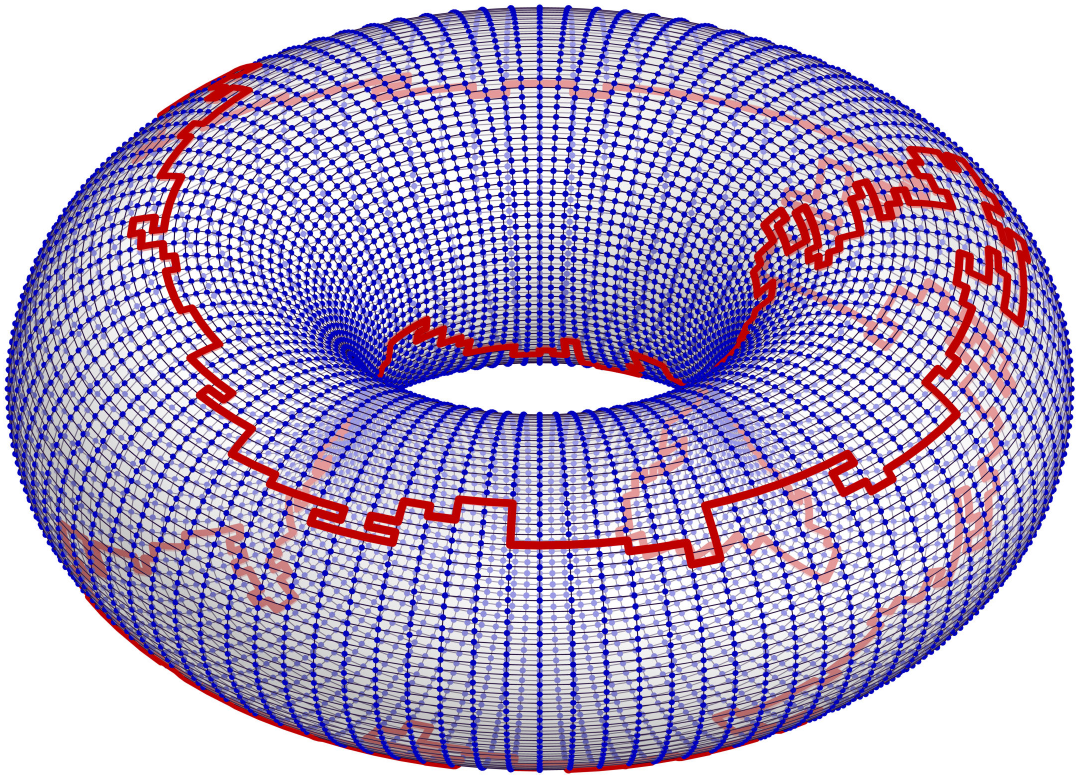}
    \caption{Example of a cycle on $ \mathbf{T}^2$ after overlaying the ground state and the maximum weight excited state as described in \eqref{eq:sym diff}. The graph type is a bipartite square lattice with $2N = 100$ and cycle length $S_{\hat{e}} = 784$.}
    \label{fig:my_label}
\end{figure}

The objective of this Chapter is to study the properties these closed curves and relate the findings to the results found in \cite{caracciolo2021criticality}. The motivation behind the study is to see how sensitive the reported critical exponents are to a different kind of excitation, namely the maximum weight excitation. In \cite{caracciolo2021criticality} strong signs of criticality and conformality were shown for the generated curves after a random link excitation. The random link excitation has a uniform measure over all the links, thus after averaging out individual randomness the underlying properties of the curves arise. In the excitation studied in this paper, the edge $\hat{e}$ with the maximum weight generates a (expected) larger rearrangement. It is indeed a sort of ``worst case scenario'' from a stability perspective, but still induced by a local perturbation. The question remains: do the curves generated after a maximum weight excitation exhibit the same kind of critical behaviour?
%it is the model that behaves and whether they belong to the same universality class.

The model was investigated for a range of system sizes: $8 \le L \le 160$ for all lattice types and the reported values are averages of $10^4 - 10^5$ instances.  Ideally, the model under study here could be ran for larger system sizes, however due lack of computational power it was infeasible.

\section{The distribution of the excitation size}

To understand the robustness of the scaling exponents reported in \cite{caracciolo2021criticality}, the same exponents are estimated for the maximum weight excitation. The maximum weight excitation selects the edge with the maximum weight and thus it is expected to create the most "damage". As a result, we expect longer curves on average with the same properties overall.

Consider the probability distribution function $P[S_{\hat{e}} > s]$ for a curve length greater than $s$. From scaling theory, the form of such probability distribution is:
\begin{equation} \label{eq:prob scale}
    P[S_{\hat{e}} > s] = s^{-\zeta} \rho\left(\frac{s^{\lambda}}{L}\right)
\end{equation}
where $\rho$ is some continuous and homogeneous scaling function that satisfies $0 < \lim_{z \rightarrow \infty} \rho (z) < \infty$. 

\begin{figure}[h]
    \centering
    \includegraphics[width =\textwidth]{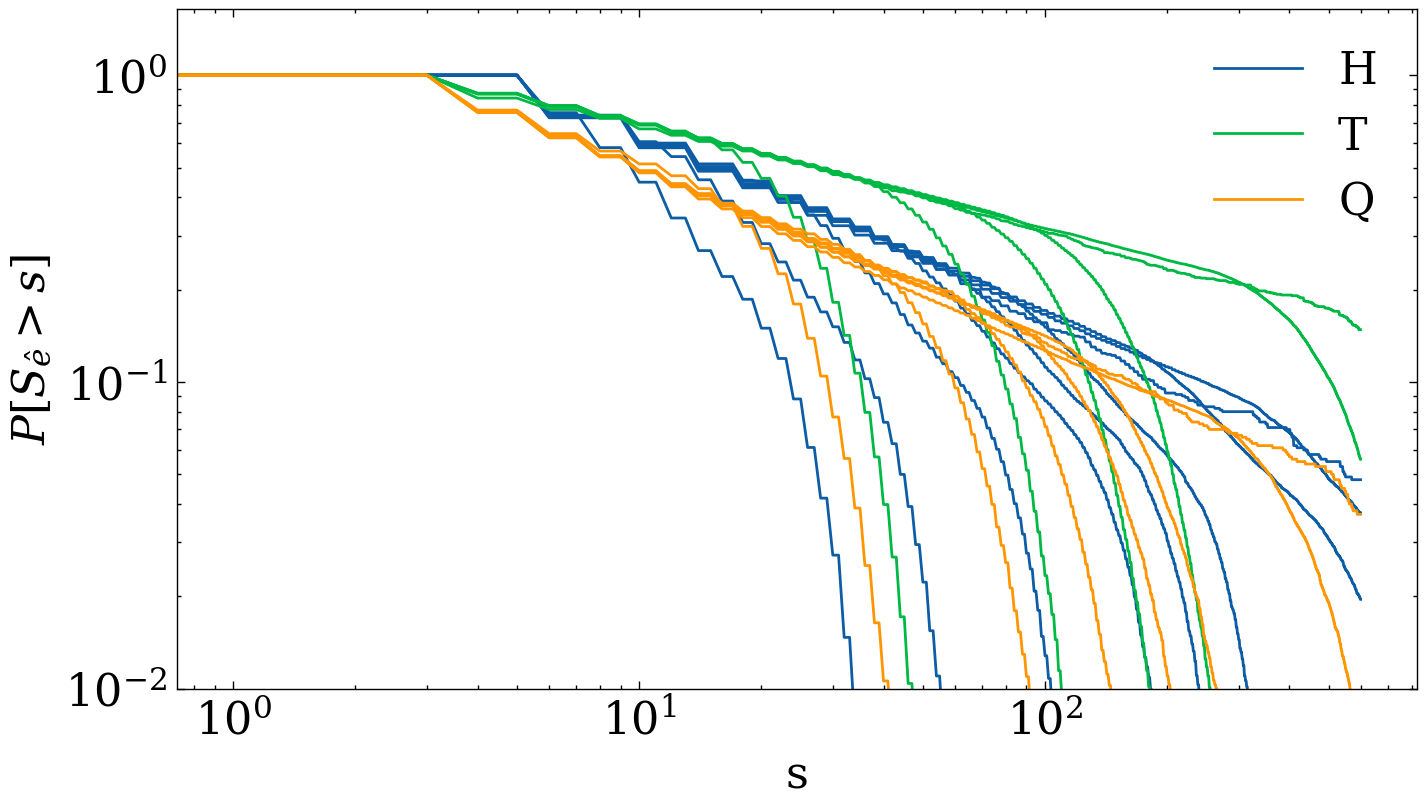}
    \caption{Cumulative distribution of $S_{\hat{e}}$ for all models. The system sizes considered are from $L=8$ to $L=256$. The sharp decays indicate strong finite size corrections. }
    \label{fig:prob fig}
\end{figure}

This scaling ansatz is numerically approximated for all models considered. In Fig. \ref{fig:prob fig} we report the numerical estimates of the previous quantity for different sizes. It is clear that a power law tail develops in all considered cases. The power law implies that local excitations induce long range rearrangements with non-zero probability in the $N \rightarrow \infty$ limit. This proves once again that the ground state solution of the random dimer model is at a critical point.  The size of the system naturally gives an upper bound to the maximum size of $S_{\hat{e}}$. Hence,  strong finite size corrections are present. The values of the exponent $\zeta$ were found by fitting a power-law tail to the data depicted in Fig. \ref{fig:prob fig}. The reported estimates for the exponents can be seen in Table \ref{Table estimates}.

The estimates for $\zeta$ from measuring the slope of the power law tail match with the reported values in \cite{caracciolo2021criticality} for all models to the first decimal point. For the $Q$ model, the values are within 1 standard deviation of the mean thus there is complete agreement. For the $H$ and $T$ models the values are underestimated, there is a disagreement from the 2.decimal point. This can be the consequence of strong finite size scaling bounds induced by the physical limit of the models. In \cite{caracciolo2021criticality} larger system sizes with many more instances were considered which explains the slight disagreement between the values. 

A known difference between the different lattice structures also holds. The $H$ and $Q$ lattices are bipartite and they belong to a different universality class than the monopartite $L$ lattice. The magnitude of the exponents in the two universality classes coincide with \cite{caracciolo2021criticality}.

\section{Estimating the fractal dimensions}
%include part about fractals in review chapter

\begin{figure}[p]
    \centering
    \includegraphics[width = \textwidth]{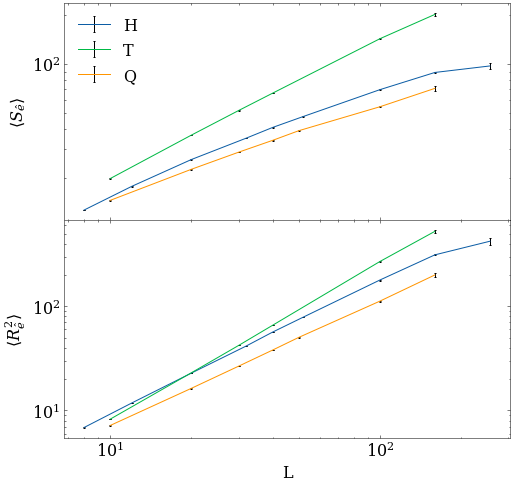}
    \caption{Scaling behaviour of $\langle S  \rangle\sim L^\alpha$ (top) and $\langle R^2 \rangle \sim L^\gamma$ (bottom) for all models studied. Points represent mean values of $ 2\times 10^3$ instances with error bars showing the statistical errors of the estimates.}
    \label{fig:alpha gamma}
\end{figure}

The fractal dimension $D_f$ of the loop quantifies the complexity in the curves as the fraction between the rate of change in detail to the rate of change in scale. It is fundamentally related to SLE and is highly relevant to describe characteristics of random curves. Assuming that the average length of the cycle scales as:
\begin{equation}
    \langle S_{\hat{e}} \rangle \sim L^\alpha
\end{equation}
which is reasonable after observing the power law relationship in Fig. \ref{fig:alpha gamma}. From, \eqref{eq:prob scale} we have
\begin{align}
    \langle S_{\hat{e}} \rangle  
    & = \int_0^\infty s \left [\frac{\partial}{\partial s}P[S_{\hat{e}} > s]\right ] ds \\
    & = \int_0^\infty s \left [\frac{\partial}{\partial s}\rho\left(\frac{s^\lambda}{L}\right)\right ] ds \\
    & = \int_0^\infty  s^{-\zeta} \rho\left(\frac{s^\lambda}{L}\right) ds
\end{align}
where in the last step we integrated by parts. Substituting in $u = \frac{s^\lambda}{L}$ we get
\begin{align}
    \langle S_{\hat{e}} \rangle  
    & = \frac{L^{\frac{1}{\lambda}}}{\lambda} \int_0^\infty (uL)^{\frac{-\zeta}{\lambda}} \rho(u) u^{\frac{1}{\lambda} - 1} du 
\end{align}
and then separating the $L$ - dependent and -independent terms we get
\begin{equation} \label{eq:alpha}
    \langle S_{\hat{e}} \rangle \propto L^{\frac{1 - \zeta}{\lambda}} \sim L^\alpha 
    \Rightarrow
    \alpha = \frac{1-\zeta}{\lambda}.
\end{equation}

To get rid of $\lambda$, we consider another quantity of interest, namely the gyration radius of a cycle, defined as
\begin{equation}
    R_{\hat{e}}^2 = \frac{1}{2 S^2_{\hat{e}}} \sum_{i,j} d^2 (r_i, r_j)
\end{equation}
where $r_i$ is the position of the ith node and $d^2(\bullet,\bullet)$ is the squared distance between the points. The sum runs over all pairs of vertices in a cycle. The gyration radius satisfies the form of a power law that is conditioned on the cycle length as:
\begin{equation}
    \langle R_{\hat{e}}^2 \rangle_{S_{\hat{e}} = s} = s^{{2}{D_f^{-1}}} g\left(\frac{s^\lambda}{L}\right) 
\end{equation}
where $g$ has the same properties as $\rho$ from \eqref{eq:prob scale}. Now, assuming the gyration radius (without conditioning) scales
\begin{equation}
     \langle R_{\hat{e}}^2 \rangle \sim L^\gamma
\end{equation}
which can be confirmed visually from Fig. \ref{fig:alpha gamma}. Inserting \eqref{eq:prob scale} we have 
%{\color{red} MISSING ARGUMENTS OF $g'$}
\begin{align}
     \langle R_{\hat{e}}^2 \rangle 
     & = \int_0^\infty \langle R_{\hat{e}}^2 \rangle_{S_{\hat{e}} = s} \left [\frac{\partial}{\partial s}P[S_{\hat{e}} > s]\right ] ds \\
     & = \int_0^\infty\left [ 2 D_f^{-1} s^{2 D_f^{-1}-1} g\left(\frac{s^\lambda}{L}\right) + s^{2 D_f^{-1}}  \frac{\lambda s^{\lambda-1}}{L} g'\left(\frac{s^\lambda}{L}\right) \right ] \left [ s^{-\zeta} \rho\left(\frac{s^\lambda}{L}\right) \right] d s
\end{align}
where we integrated by parts. Substituting in $u=\frac{s^\lambda}{L}$ we get
\begin{align}
    \langle R_{\hat{e}}^2 \rangle = \int_0^\infty
    \left [ (uL)^{\frac{2 D_f^{-1} -1}{\lambda}} g(u) + (uL)^{\frac{2D_f^{-1}}{\lambda}} \frac{\lambda (uL)^{\frac{\lambda -1}{\lambda}}}{L}  g'(u) \right ]
    \left [ (uL)^{\frac{-\zeta}{\lambda}} \rho(u) L^{\frac{1}{\lambda}} \frac{1}{\lambda} u^{\frac{1}{\lambda} - 1} \right ] du 
\end{align}
 and separating the $L$-dependent and $L$-independent terms we get
\begin{equation}\label{eq:gamma}
    \langle R_{\hat{e}}^2 \rangle \propto L^{\frac{2D_f^{-1} - \zeta}{\lambda}} \sim L^\gamma
    \Rightarrow
    \gamma = \frac{2 D_f^{-1} - \zeta}{\lambda}.
\end{equation}

\begin{table}[p]
    \centering
    \begin{tabular}{llll} 
        \hline\hline
                                     & \multicolumn{1}{c}{H} & \multicolumn{1}{c}{Q} & \multicolumn{1}{c}{T}  \\ 
        \hline
        $\langle \Delta E_{\hat{e}} \rangle $     & 1.283(4)                & 0.759(7)               & 0.454(2)                  \\
        $\alpha$                     & 0.576(11)               & 0.591(17)                 & 0.867(3)                  \\
        $\gamma$                     & 1.290(3)                 & 1.208(18)                & 1.507(7)                  \\
        $\zeta$ (from fit)           & 0.579(1)                 & 0.602(2)                 & 0.341(1)                  \\
        $\zeta$                      & 0.552(18)                & 0.573(53)                 & 0.363(19)                 \\
        $D_f$                        & 1.287(14)               & 1.383(35)                 & 1.360(11)                  \\
        $\kappa$                     & 2.093(7)                 & 2.035(4)                 & 2.168(4)                  \\
        $D_f (1 + \frac{\kappa}{8})$ & 1.262(1)                 & 1.254(1)                 & 1.271(1)                  \\
        \hline\hline
    \end{tabular}
    \caption{Average ground state cost, average change in the energy and scaling exponent estimates. Values in brackets indicate statistical errors.}
    \label{Table estimates}
\end{table}

\begin{table}[p]
    \centering
    \begin{tabular}{llll} 
        \hline\hline
        & \multicolumn{1}{c}{H} & \multicolumn{1}{c}{Q} & \multicolumn{1}{c}{T}  \\ 
        \hline
        $\langle N^{-1} E[D^*] \rangle $                                       & 0.703(1) & 0.355(1) & 0.529(1)  \\
        $\langle \Delta E_{\hat{e}} \rangle $                                      & 1.115(1) & 0.655(1) & 0.380(1)  \\
        $\alpha$                 & 0.508(2) & 0.506(1) & 0.827(1)  \\
        $\gamma $      & 1.259(1) & 1.252(2) & 1.554(1)  \\
        $\omega$ & 0.74(2)  & 0.65(2)  & 1.00(2)   \\
        $\zeta$                                          & 0.593(1) & 0.597(1) & 0.350(1)  \\
        $\zeta$ (from fit)                                 & 0.593(1) & 0.595(1) & 0.354(1)  \\
        $D_f$                                                          & 1.250(2) & 1.253(2) & 1.273(1)  \\
        $\kappa$                                         & 2.034(6) & 2.003(7) & 2.181(4)  \\
        $D_f (1 + \frac{\kappa}{8})$ & 1.254(1) & 1.250(1) & 1.273(1)  \\
        \hline\hline
    \end{tabular}
    \caption{Copy of parameter estimates from \cite{caracciolo2021criticality}. Table inserted here to ease the comparison of values.}
\end{table}

Combining the expressions \eqref{eq:alpha} and \eqref{eq:gamma} on the exponents $\alpha$, $\gamma$ and $\zeta$, we can get rid of $\lambda$ and the fractal dimension $D_f$ can be expressed as
\begin{align}
    D_f & = 2- \gamma + \alpha \\
    \zeta & = \frac{2 - \gamma}{D_f} 
\end{align}
which allows the calculation of the fractal dimension from the exponents $\alpha$ and $\gamma$, which in turn are measurable through accurate numerical simulations. 

The measurement of these two exponents was accomplished by fitting $\ln\langle S\rangle$ and $\ln\langle R^2\rangle$ against $\ln L$ with a function of the form
\begin{equation}
     f(x) = \omega^{(1)} +   \omega x + \omega^{(2)} e^{- \omega^{(3)} x}
\end{equation} where $\omega$, $\omega^{(1)}$, $\omega^{(2)}$ and $\omega^{(3)}$ are free parameters. The additional exponential term is needed to account for the finite-size corrections to the asymptotic regime. %That means that the lenght of cycles will never extend freely to infinity, they will always be constrained by the system size $L$. 
A more refined way to extrapolate the exponents in the $L\to\infty$ limit is the method of ratios introduced in \cite{caracciolo1995extrapolating} and adopted in \cite{caracciolo2021criticality}. We are however unable to perform that technique due to lack of reliable data: instead, an additional term with the argument and two free parameters $\omega^{(2)}$ and $\omega^{(3)}$ has been added to the fitting function $f(x)$. 

Hence, to estimate $\alpha$ for the scaling of the average cycle length $ \langle S_{\hat{e}} \rangle \sim L^\alpha$, $f(x)$ is fitted as:
\begin{equation}
    \ln \langle S_{\hat{e}} \rangle  = \alpha^{(1)} +   \alpha \ln L + \alpha^{(2)} L^{- \alpha^{(3)}}.
\end{equation}
%with free parameters $\alpha^{(1)}$, $\alpha^{(2)}$ and $\alpha^{(3)}$. 
To measure $\gamma$ for the proposed scaling of the gyration radius $\langle R_{\hat{e}}^2 \rangle \sim L^\gamma $, $f(x)$ is fitted as:
\begin{equation}
    \ln \langle R_{\hat{e}}^2 \rangle = \gamma^{(1)} +   \gamma \ln L + \gamma^{(2)} L^{- \gamma^{(3)}}
\end{equation}
The function fits are implemented through SciPy's \cite{virtanen2020scipy} curve fit method. The parameter value estimates are found in Table \ref{Table estimates}. 

%gamma is sometimes more sometimes less indicates it is around the true value
The reported values for $\gamma$ are of the same precision as the values for $\zeta$. They match up to the 2. decimal point for all models, although there is some discrepancy. This difference is attributed to the strong finite size corrections of the system and that different statistical methods were used in this report. The separation between lattice types is clear for the values of $\gamma$, although not as clearly as for other parameters.

%alpha is always more
On the other hand, there is some disagreement between the values of $\alpha$ found here and in \cite{caracciolo2021criticality}. The estimates of $\alpha$, which describes the scaling of the average length of the cycles as a function of the system size tends to be larger on the second decimal point in all cases for the maximum weight excitation. On one side, we expect the loops obtained by means of this excitation strategy to be larger than the one obtained in \cite{caracciolo2021criticality}. The average cost of the loops $\langle \Delta E_{\hat{e}} \rangle$ are also higher for all models compared to \cite{caracciolo2021criticality} which confirms that the loops are indeed on average longer. However, we cannot exclude that the exponent $\alpha$ is identical to the one obtained using a random edge excitation, and more precise numerical estimates are needed to clarify this point. The precision is however sufficient to observe the separation of values for the different lattice types in this case too. 

The fact we estimate a larger $\alpha$ with respect to \cite{caracciolo2021criticality} induces in turn a different, larger value for the fractal dimension from the relationship $D_f = 2 -\gamma + \alpha$ for all models. For example for the max weight excitation $Q$ model $D_f = 1.383 \pm 0.036$ while for the random excitation $Q$ model $D_f = 1.254 \pm 0.002$. As we mentioned above, we think that the discrepancy is caused by the fact that reported values are affected by corrections due to the finite system sizes and larger system sizes ($L > 500$) need to be simulated.
%or (ii) the statistical methods used to estimate the values do not properly generalise to the limit $L \rightarrow \infty$. {\color{red} WHAT DO YOU MEAN HERE?}

Using the relation $\zeta = \frac{2-\gamma}{D_f}$, the values calculated for $\zeta$ are in agreement with the measured values from the data to the second decimal point. This fact comforts the validity of the underlying scaling relation between the parameters $\alpha$, $\gamma$ and $\zeta$ also in this case.

\section{Winding angle variance}

The estimated values for the scaling of the average cycle length $ \langle S_{\hat{e}} \rangle$ and gyration radius $\langle  R^2_{\hat{e}} \rangle$ indicate that the system is critical. Hence, the presence of a stronger form of symmetry is tested: conformal invariance. Following the methodology in \cite{caracciolo2021criticality}, the average winding angle variance $\langle \theta^2 \rangle$ is calculated. %it has been shown that theta is related to kappa bla bla...
As seen in \cite{Corberi_2019,wieland2003winding}, if the ensemble of $\text{SLE}_\kappa$ curves is considered, $\langle \theta^2 \rangle$ is directly connected to the parameter $\kappa$ through the following equation:
\begin{equation}
    \langle \theta^2 \rangle = a + \frac{\kappa}{4} \ln L
\end{equation}
when $\theta$ is defined as the difference in angles between subsequent edges: $\theta(e + 1) = \theta(e) + \mathrm{angle}(e, e+1)$ expressed in radians. For example for the square lattice the angle function takes only three possible values  $\{\frac{\pi}{2}, -\frac{\pi}{2}, \pi\}$. If we restrict ourselves to curves winding around the torus (so the longer ones), then $\sum_{e\in S_{\hat{e}}}\theta(e) = 0$ since $S_{\hat{e}}$ is a closed loop. The average variance of the winding angle is
\begin{equation}
    \langle \theta^2 \rangle = \frac{1}{S_{\hat{e}}} \sum_{e \in S_{\hat{e}}} \theta^2(e).
\end{equation}
In this case fitting a function of the form
\begin{equation}
    \langle \theta^2 \rangle = a + \kappa' \ln L
\end{equation}
such that $\kappa = 4 \kappa' $ is used to estimate the parameter $\kappa$, with a free parameter $a$. Furthermore, the values for the known connection between $\text{SLE}_\kappa$ and the fractal dimension $D_f$ is reported in Table \ref{Table estimates}.
\begin{equation}
    D_f = \min\left(1 + \frac{\kappa}{8}, 2\right) 
\end{equation}

\begin{figure}[h]
    \centering
    \includegraphics[width = \textwidth]{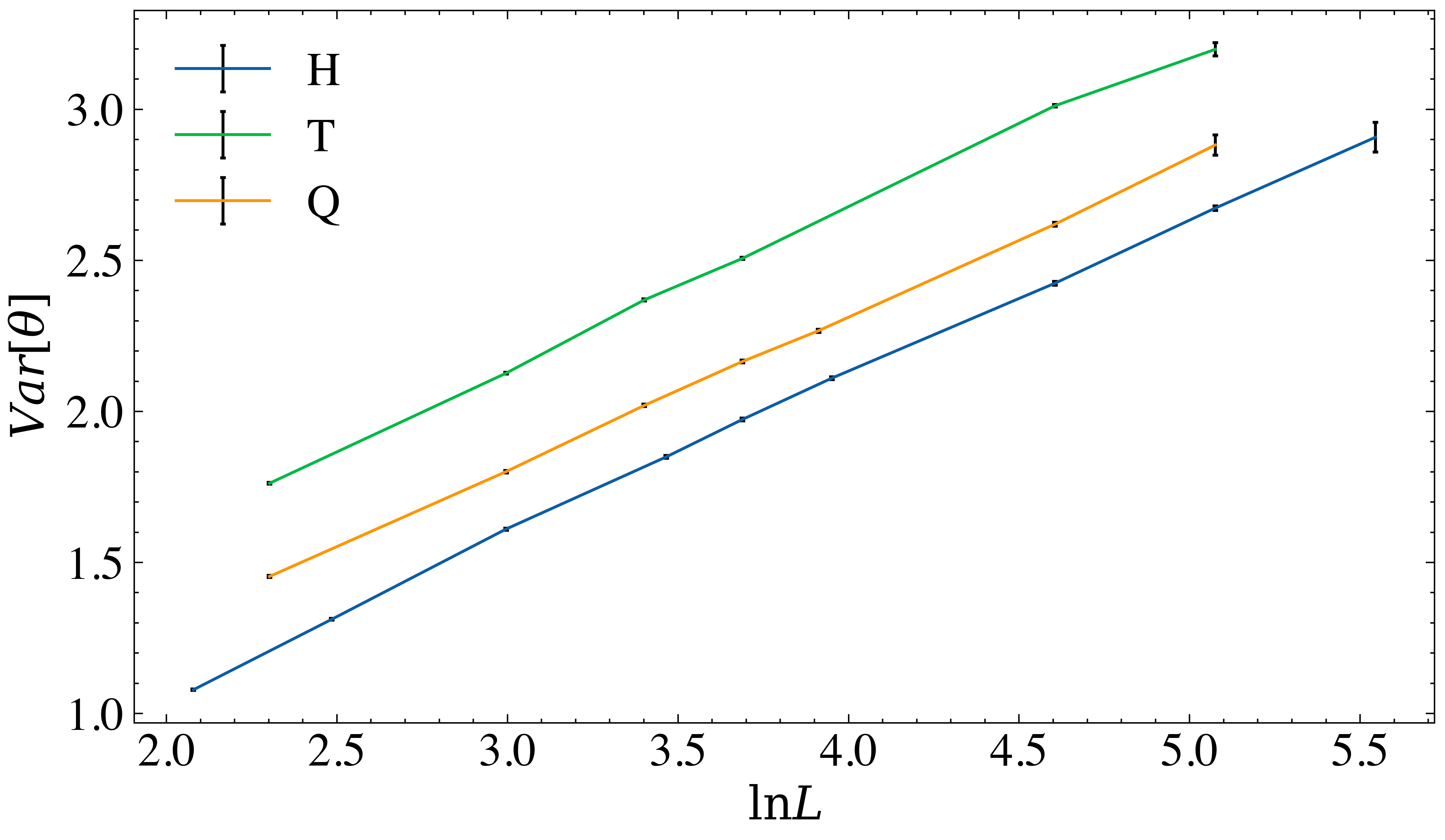}
    \caption{Scaling relation of the variance of the winding angle $\theta$ of the cycles for all models. Points represent averages of $2 \times 10^3$ instances with error bars showing the statistical errors of the estimates.}
    \label{fig:kappa}
\end{figure}
In all cases, the estimated values for $\kappa$ are in excellent agreement with values reported in \cite{caracciolo2021criticality}. For all models, the estimates match to the 3. decimal point. Consequently, the fractal dimension of the cycles $D_f = \min(2,1+\frac{\kappa}{8})$ corresponds well to the ones reported in \cite{caracciolo2021criticality}.

%temporary section
\section{Conclusions and perspectives}

In this chapter, we tested the sensitivity of critical exponents reported in \cite{caracciolo2021criticality} to different excitation strategies on three different lattices. Instead of the random link excitation, the maximum weight excitation was applied, which forbids the edge with the largest weight: we expect that this excitation causes a particularly large excitation of the ground state. The same properties of the curves generated by overlaying the two solutions were explored as in \cite{caracciolo2021criticality}. 

We find that the average length of the curves $\langle S_{\hat{e}} \rangle\sim L^\alpha$, with $L$ size of the system, after the maximum weight excitation. We found a value of $\alpha$ slightly larger than the one found exciting the system by cutting a randomly chosen edge in the solution. As expected, the average cost of the curves $\langle \Delta E_{\hat{e}} \rangle$ is also larger. Consequently, the fractal dimension of the curves is larger than the one obtained using a random strategy. We believe that this first approach overestimate the fractal dimension of the curve and that the larger values of $\alpha$ are due to the presence of large finite-size effects. This is supported by the fact that a different estimation strategy, related to the scaling of the variance of the winding angles $\langle \theta^2 \rangle$, gives instead values of the fractal dimension fully compatible with \cite{caracciolo2021criticality}. To confirm this hypothesis, a more refined analysis (e.g., using the method of ratios) has to be performed on larger system sizes, that however requires larger computational time. %In \cite{caracciolo2021criticality} the random excitation can be viewed as selecting an edge randomly, with a uniform measure over all edges while here the maximum weight edge has probability 1 of being selected. The random excitation averages over over the weight distribution.  Therefore, the random link excitation is sensitive on the sampling of the weight distribution. In this case, the $\alpha$ parameter could include an additional correction term: $\alpha' = \alpha - c $ to account for selecting the edge with the largest weight, where $c$ is a model dependent parameter. 

Overall, the parameters show similar values for the maximum weight excitation as for the random weight excitation. In particular, the values of $\kappa$ calculated from the winding angle variance $\langle \theta^2 \rangle$, which has been used as a tool for a conformality test \cite{wieland2003winding}, confirm that the cycles on bipartite graphs $H$ and $Q$ have dimension compatible with the one of the loop-erased self avoiding random walk, which has $D_f = \frac{5}{4}$ and is $\text{SLE}_2$. The $T$ model, which is monopartite exhibits excitation dimension in line with the one of 2d spin glass domain walls with $D_f = 1.271$, which is believed to be an $\text{SLE}_\kappa$ with $\kappa \approx 2.1$.

\chapter{Near-optimal configurations in random dimer coverings} \label{ch.4}

The stability of the ground state solution of the random dimer model is further investigated in this chapter. The excitations on the random dimer model considered so far in Ch. \ref{chap: max weight} and in \cite{caracciolo2021criticality} are local excitations: an edge is deleted which always results in some kind of rearrangement (that we saw that it is long-ranged with finite probability). As a result, a link excitation is not suitable to understand the stability of the lowest energy configuration. To understand the energy landscape of the dimer model in close proximity to the lowest energy configuration, a continuous excitation is needed (or, from a practical point of view, a discrete one with tunable step sizes).

The aim of this chapter is to apply a \textit{smooth} perturbation to the ground state solution of the random dimer model. The perturbation is driven by a scalar $\epsilon$ such that weights on edges that are in the ground state solution are penalised by $\epsilon$. By varying the parameter $\epsilon$ it allows a controlled near-continuous transition between the lowest energy configuration and excited states. The method draws from \cite{pagnani2003near} where the same kind of excitation was considered on a complete graph, which in turn extends the ideas of \cite{aldous2003scaling}. 

In \cite{aldous2003scaling} various combinatorial optimisation problems defined on random domains, for example the traveling salesman problem and the minimum spanning tree problem were considered. It is conjectured that optimisation problems can be classified into a small number of universality classes according to the ground state solution's dependence on small excitations. To which universality class a given problem belongs to is given by its relative difficulty of finding the optimal solution. For example, the minimum spanning tree is considered an algorithmically easy problem and thus near-optimal configurations scale differently than in the traveling salesman problem (TSP) or the minimum-weight perfect matching, which are algorithmically harder. In \cite{pagnani2003near} this conjecture is investigated further for the minimum weight perfect matching on a complete graph. Their numerical results are in agreement with \cite{aldous2003scaling} and hint towards the underlying universality. 

The aim of this chapter is to investigate this conjecture for the random dimer model by computing the same scaling exponents. The random dimer model is a minimum weight matchings on a lattice, hence if the conjecture holds we expect similar scaling behaviour of near-optimal configurations as on complete graphs \cite{pagnani2003near}. This is because if the proposed conjecture in \cite{aldous2003scaling} is true, the determining force for the scaling behaviour is algorithmic complexity, in which case the underlying graph structure is less important. 

%say about the difficulty of the problems belong to different categories

%describe what versions of the model is done in those papers

\section{The $\epsilon$-excitation}

Let us consider again the same setup for the random dimer model on a weighted graph $G(V,E)$ as described in Sec. \ref{RDM}. The edge weights are drawn from an exponential distribution $q(w) = e^{-w}$ and the ground state $D^*$ is defined as the dimer covering that minimises the sum of weights: $E[D^*] = \min_D E[D]$. In this case, for a given set of edge weights $w_e$ the perturbation of the model is introduced by adding an additional $\epsilon$ weight to the edges found in the ground state solution of the model $ e \in D^* $. Introducing an indicator variable $n_e$ for edges in the ground state solution, the new weights are given by
\begin{equation}\label{weights}
    w_e^\epsilon = w_e + n_e\epsilon
\end{equation}
where $\epsilon$ is global parameter of the model. All edges in the ground state dimer covering ${e} \in D^*$ are penalized identically while edges that are not in the ground state solution $e \notin D^*$ are not penalized. Then, the algorithm for finding the maximum weight perfect matching is run again, giving the $\epsilon$-ground state or perturbed solution $D^*_{{\epsilon}}$. In this case, $\epsilon$ gives an additional degree of freedom to the model and allows a smoother excitation than cutting a link instantly. As $\epsilon$ is increased, the ground state will be penalised more heavily so that at a certain (instance dependent) value $\epsilon^*$ the former ground state is not optimal anymore. 

The $\epsilon$-ground state solution is characterized by the indicator variable $n_e^\epsilon$ for edges that are in the optimal configuration. We define the difference in the cost between the ground state $D^*$ and $\epsilon$-ground state $D^*_{{\epsilon}}$ is
\begin{equation}\label{alpha scale rel}
    \Delta E_{\epsilon} = \sum_e (n_e^\epsilon - n_e)w_e
\end{equation}
where the sum runs over all edges. 
%The difference in cost which is equal to zero \textit{iff} $\epsilon = 0$.Note, that even though a non-zero $\epsilon$ weight might not induce any rearrangement, it will always strictly increase the cost of the ground state. 
Another quantity of interest is the distance between the $\epsilon$-ground state and the original ground state, which is defined as:
\begin{equation}
    d_{{\epsilon}} = 1 - l_{{\epsilon}}
\end{equation}
Here, the overlap $l_{\Tilde{\epsilon}}$ between the two coverings is given by:
\begin{equation}
    l_{{\epsilon}} = \frac{1}{N}\sum_{e} n_e^\epsilon n_e .
\end{equation}
where the sum runs over all edges of the graph.

Various lattice types, i.e., honeycomb(H), square (Q) and triangular (T) (see Fig. \ref{fig:lattice types}) and system sizes $L=\{20,50,100\}$ are considered as well as the perturbation size is varied $\epsilon \in [0.01, 0.9]$\footnote{The values $\epsilon$ are not uniform. Higher density towards $0$.}. For each setting, averages of $10^3-10^4$ simulation instances are collected. To find the ground state of the model, the blossom algorithm \cite{kolmogorov2009blossom} is used, which finds the optimal configuration with time complexity $O(EV \log V)$. For an overview of the algorithm see Sec. \ref{Edmonds}. In \cite{pagnani2003near} the mean-field version of the problem (i.e., $G$ is assumed to be a complete graph) is studied and the cavity method is used to find optimal configurations. As they showed, the prediction from the cavity method perfectly coincides with the blossom algorithm results. %, thus we assume that no differences in the results are due to the different methods used. 

\section{Scaling exponents}

The $\epsilon$-excitation is numerically studied with the aim to better understand the scaling behaviour of near-optimal configurations. If we take the ground state solution of a realisation of the random dimer model, we are interested in how far in distance the excited ground states are from it for a given $\epsilon$ value. Furthermore, we are interested in how the normalised change in energy $\frac{\Delta E_{\epsilon}}{N}$ scales as a function of the distance $d_{\epsilon}$.

After \cite{aldous2003scaling} and \cite{pagnani2003near}, suppose the scaling relation for the distance as a function of the perturbation:
\begin{equation}
    d \sim \epsilon^\beta
\end{equation}
in close proximity to the ground state solution. The exponent $\beta$ describes the expected distance for an excitation of size $\epsilon$. Suppose also a power-law relation for average energy change as a function of the distance to the ground state
\begin{equation}
    \frac{\langle \Delta E_{\epsilon}\rangle}{N} \sim d^\tau.
\end{equation}

Observe, that the scaling exponents $\beta$ and $\tau$ are not independent. From \eqref{alpha scale rel}, we see that $\Delta E_\epsilon \sim \frac{N}{2}d \epsilon$. This is because there are $Nd/2$ nonzero terms in the sum. The $\epsilon$ term comes from the fact that we expect the new edge weights to be larger than the old weights by an amount that is order of the perturbation $\epsilon$. Thus the expected scaling is
\begin{equation}\label{alphabeta}
    \Delta E_\epsilon \sim \epsilon^{\beta + 1} \sim d^{\frac{\beta + 1}{\beta}}\Longleftrightarrow\tau = \frac{\beta+1}{\beta}.
\end{equation}

\begin{figure}[p]
    \centering
    \includegraphics[width=\textwidth]{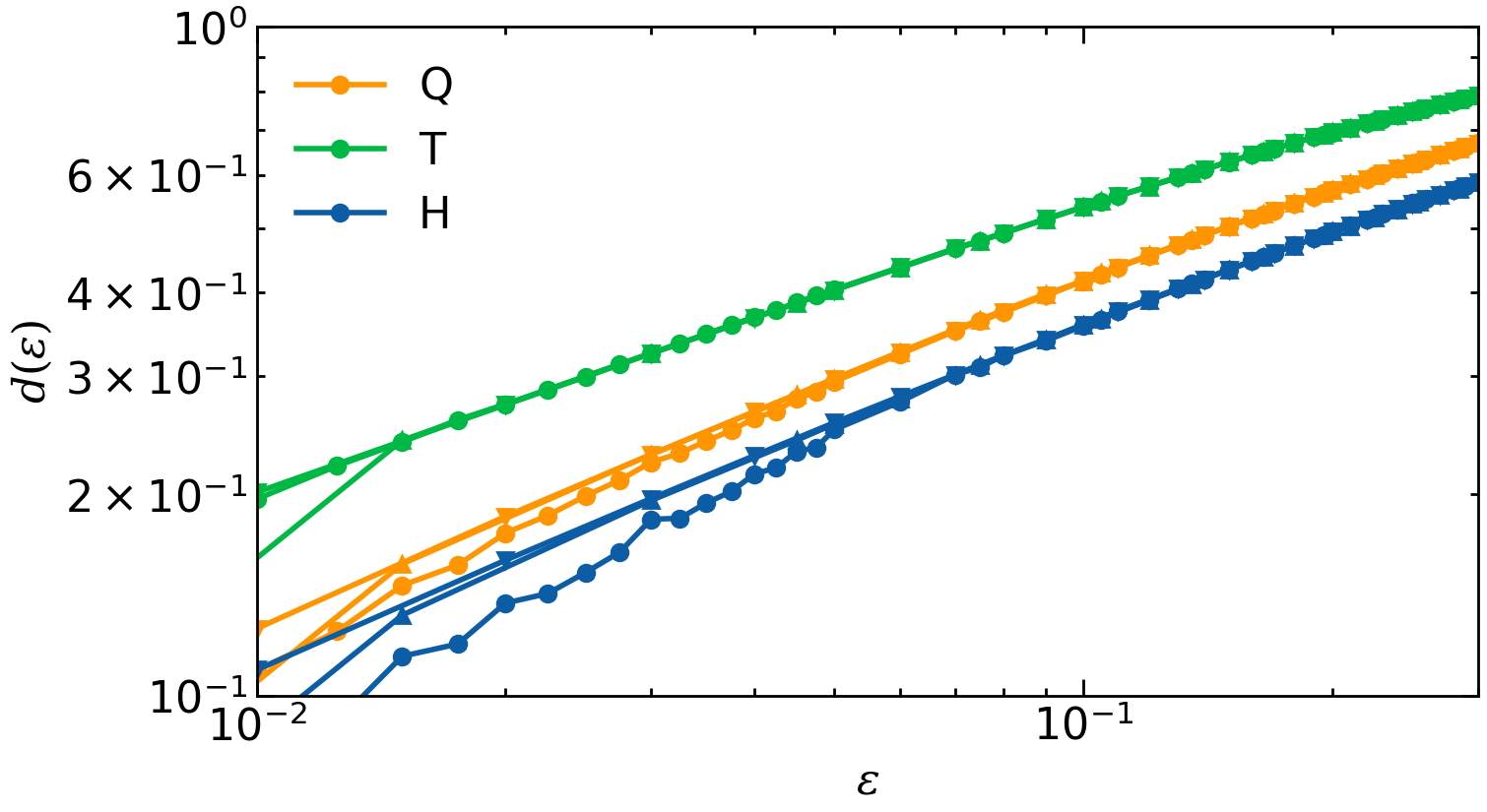}
    \caption{Scaling of distance $d$ as a function of the perturbation size $\epsilon$. Different system sizes are indicated by different shapes: $L=20$ circular, $L=50$ inverse triangular and $L=100$ triangular dot. Statistical errors are invisible.}
    \label{fig:beta scaling}
\end{figure}

\begin{figure}[p]
    \centering
    \includegraphics[width=\textwidth]{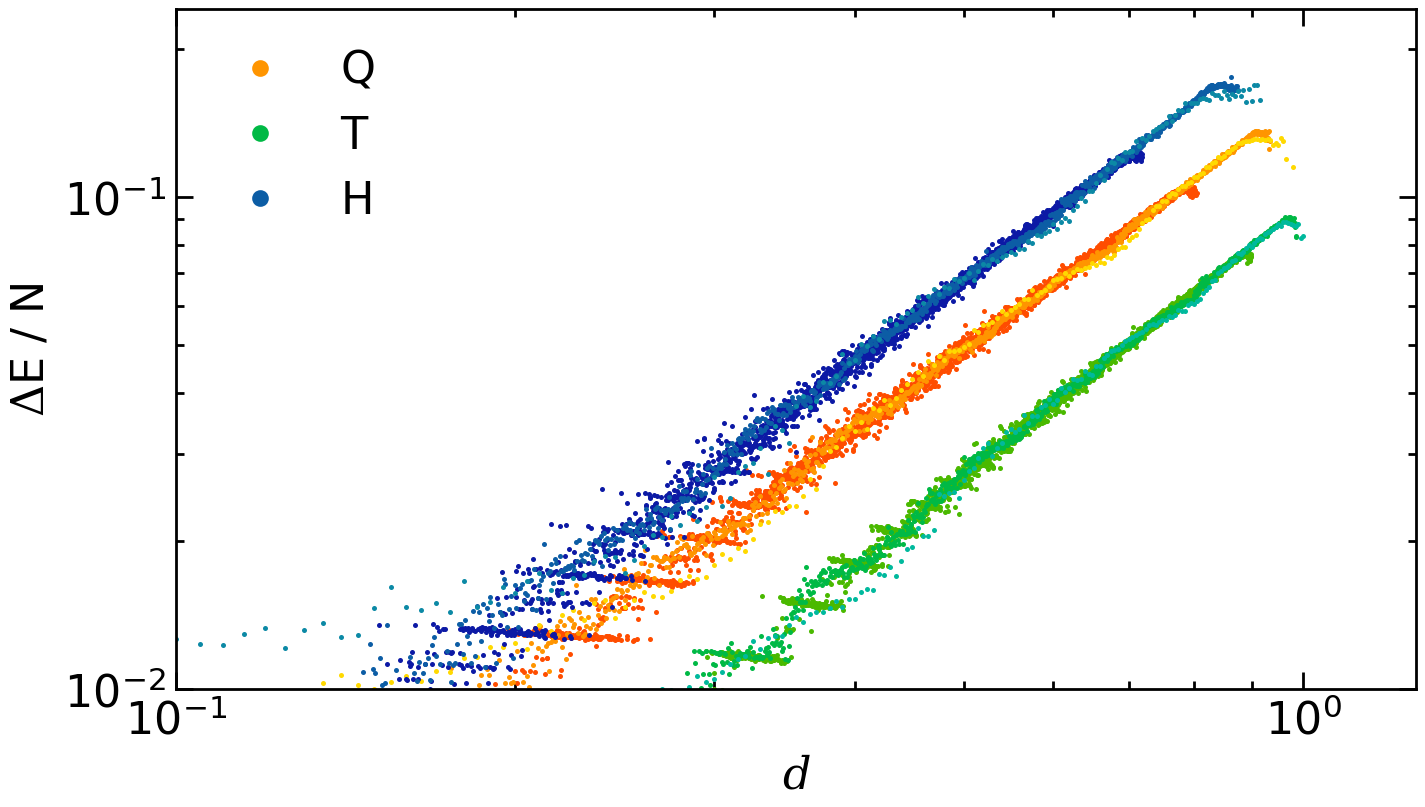}
    \caption{Scaling of the normalised average energy difference with respect to distance $d$. Different system sizes are indicated by different shades of the color: $L=20$ light, $L=50$ medium and $L=100$ dark shade. The visible "stairs" in the graph are due to discrete steps in $\epsilon$. }
    \label{fig:tau scaling}
\end{figure}

To estimate $\beta$ and $\tau$ we look at Fig. \ref{fig:beta scaling} and Fig. \ref{fig:tau scaling} respectively and fit a function of the form to the data
\begin{equation}\label{fit func}
    f(x) = \omega^{(1)} + \omega x
\end{equation}
where $\omega^{(1)}$ is a constant and $\omega$ gives the slope. The values are reported in Table \ref{estimates 2}, with errors in brackets.

\begin{table}[b]
    \centering
    \begin{tabular}{llll} 
    \hline\hline
           & \multicolumn{1}{c}{H}        & \multicolumn{1}{c}{Q}        & \multicolumn{1}{c}{T}       \\ 
    \hline
    $\beta$    & 0.528(11) & 0.535(15) & 0.445(10)  \\
    $\tau$ & 3.001(68) & 2.721(44) &  3.119(64)  \\
    $\frac{1 + \beta}{\beta}$ & 2.893(39) & 2.869 (52) & 3.247 (50) \\
    \hline\hline
    \end{tabular}
    \caption{Estimated values of scaling exponents $\beta$ and $\tau$ for honeycomb (H), square (Q) and triangular (T) models. The numbers in brackets indicate statistical errors.}
    \label{estimates 2}
\end{table}

We see that the numerical values of $\beta$ are scattered around $0.5$, which matches the estimates found in \cite{pagnani2003near}. Furthermore, the values of $\tau$ from the fit are spread around $3$, which hint that the models belongs to the same universality class. Plugging in the values of $\beta$ to \eqref{alphabeta}, which gives an alternative way to compute $\tau$ we see that scaling exponent for the $H$ model is $2.893 \pm 0.033$, for the $Q$ model is $2.869 \pm 0.045$ and for the $T$ model it is $3.247 \pm 0.030$. Although the values of $\tau$ and $\frac{1+\beta}{\beta}$ do not match identically, we believe they approach the same value for larger $N$. Finite size corrections of the system are present in the models, which distorts the reported values. 

Our estimates for the scaling exponents $\beta$ and $\tau$ support the conjecture proposed in \cite{aldous2003scaling}, where the scaling of the travelling salesman problem was studied. They found that the energy of near-optimal configurations scale $\Delta E \sim d^{3\pm 0.024}$ across dimensions $d=2, 3, 4$ in both the Euclidean matching problem and the Euclidean TSP. With \textit{Euclidean} it is meant that the problem is considered on a graph whose vertices are associated to random points in the Euclidean domain and the weights are the Euclidean distances between the endpoints. Although the TSP is a different problem, it is heuristically related to the minimum weight matchings. When all edges are deleted that are in the ground state solution of the minimum weight matching, and then the new optimal solution is overlayed with the ground state solution, we find a sub-optimal solution of the traveling salesman problem. The above found scaling exponents hint that the two problems belong to the same universality class as far as the value of $\beta$ is concerned. Thus, the conjecture proposed in \cite{aldous2003scaling} seems to hold for the random dimer model too.

\section{Conclusion and perspectives}

In this chapter we investigate the scaling of near-optimal configurations of the random dimer model on three types of lattices in 2 dimensions. The conjecture put forward in \cite{aldous2003scaling} that combinatorial optimisation problems belong to a small number of universality classes based on a solutions' dependence on the ground state excitation is explored. Following the method in \cite{pagnani2003near}, we generate near-optimal configurations by applying the $\epsilon$-coupling method: the edges in the ground state solution are uniformly penalized by adding an $\epsilon$ weight.

We measure two exponents that describe how the relationship between the ground state solution and the $\epsilon$-ground state: (i) $\beta$, which is the scaling exponent of the distance $d$ as a function of excitation size $\epsilon$: $d \sim \epsilon^\beta$ and (ii) the $\tau$, which describes how the average energy difference scales as a function of the distance: ${\langle \Delta E_{\epsilon}\rangle}/{N} \sim d^\tau.$ An additional, non-trivial relationship is tested between the exponents $\tau = (\beta + 1)/\beta$.

The reported exponents are compared to the ones in \cite{aldous2003scaling} and \cite{pagnani2003near}, although not precise, but agreement is found. Even though in \cite{pagnani2003near} a complete graph is studied, we as well find that $\beta$ is scattered around $0.5$ with a relatively large statistical errors, which hint towards the universality of minimum weight matchings, regardless of the graph structure. Our estimates also indicate the proposed cubic scaling of the energy difference $\tau$, which is also confirmed in \cite{pagnani2003near}. Hence, the computed exponents hint towards the universality of near-optimal configurations in combinatorial optimisation problems.

\printbibliography

\end{document}